%% file: ms.tex
\documentclass[]{aastex631}
\usepackage{CJK}
\usepackage{graphicx}
\usepackage{tabularx}
\usepackage{booktabs}
\usepackage{multirow}
\usepackage{makecell}
\usepackage{rotating}
\usepackage{subfigure}
\usepackage{gensymb}
\usepackage{placeins}  % 添加这个包以支持 \FloatBarrier 命令
\usepackage{etoolbox}

\makeatletter
\patchcmd{\@outputpage@head}{\maxdeadcycles=10}{\maxdeadcycles=100}{}{}
\makeatother

\shorttitle{Search for the young water fountain candidates}
\shortauthors{Xie et al.}
\graphicspath{{./}{figures/}}

\begin{document}
\begin{CJK*}{UTF8}{gkai}
\title{Infrared SED Modeling of Velocity-Excess Maser Sources: Identifying Incipient Water-Fountain Candidates}

\correspondingauthor{Jun-ichi Nakashima}
\email{junichin@mail.sysu.edu.cu, nakashima.junichi@gmail.com}

\author[0000-0003-1015-2967]{Jia-Yong Xie (谢嘉泳)}
\affiliation{School of Physics and Astronomy, Sun Yat-sen University, Tang Jia Wan, Zhuhai, 519082, P. R. China}

\author[0000-0003-3324-9462]{Jun-ichi Nakashima(中岛淳一)}
\affiliation{School of Physics and Astronomy, Sun Yat-sen University, Tang Jia Wan, Zhuhai, 519082, P. R. China}
\affiliation{CSST Science Center for the Guangdong-Hong Kong-Macau Greater Bay Area, \\Sun Yat-Sen University, 2 Duxue Road, Zhuhai 519082, Guangdong Province, PR China}

\author[0000-0002-1086-7922]{Yong Zhang (张泳)}
\affiliation{School of Physics and Astronomy, Sun Yat-sen University, Tang Jia Wan, Zhuhai, 519082, P. R. China}
\affiliation{CSST Science Center for the Guangdong-Hong Kong-Macau Greater Bay Area, \\Sun Yat-Sen University, 2 Duxue Road, Zhuhai 519082, Guangdong Province, PR China}

\begin{abstract}
We investigated whether “velocity excess” in circumstellar maser lines can diagnose the earliest evolutionary phases of Water Fountains (WFs). Here we define “velocity excess” as maser emission (e.g., H$_2$O 22.235 GHz or OH 1665/1667 MHz) detected at velocities outside the velocity range of the OH 1612 MHz line, which traces the terminal expansion velocity of a spherical circumstellar envelope (CSE). Such excess velocities serve as an indicator of gas motions deviating from spherical expansion and may signify the onset of asymmetric outflows. Based on recent studies \citep{2024ApJS..270...13F,2025ApJ...978..114X}, we analyzed 17 sources showing velocity excess and fitted their infrared spectral energy distributions (SEDs) with the one-dimensional radiative transfer code \textsc{DUSTY}. Seven sources are well reproduced, implying outer CSEs that remain nearly spherical despite inner asymmetries. Among these, five exhibit single-peaked, AGB-like SEDs and two show double-peaked, post-AGB-like profiles. IRAS variability indices and NEOWISE-R light curves reveal pulsations ($\sim$600–--1000 days) in three sources, supporting their AGB classification. Considering the magnitude of the velocity excess, two objects—IRAS 19229+1708 and IRAS 19052+0922—may represent the earliest or incipient WF phase, in which asymmetric outflows are beginning to emerge within otherwise spherical envelopes. These results support a morphological sequence in which bipolar jets and tori arise first in the central regions while the outer CSE remains spherical, and they show that selecting WF candidates via velocity excess effectively identifies objects at the onset of jet formation and early morphological transformation.
\end{abstract}


\section{Introduction} \label{sec: intro}

Maser lines arising from molecular species such as hydroxyl (OH), water (H$_2$O), and silicon monoxide (SiO) are commonly observed in the circumstellar envelopes (CSEs) of oxygen-rich asymptotic giant branch (AGB) stars. The physical conditions required for maser emission, such as density and excitation temperature, differ depending on the molecular species and transitions. Therefore, masers from different molecules and transitions typically occur in distinct regions within the CSE. Observations with the IRAM Plateau de Bure Interferometer have shown that approximately 70\% of AGB stars possess nearly spherically symmetric envelopes \citep{1998A&AS..130....1N}. In these spherically symmetric CSEs, masers from different molecular species tend to be distributed in spherical, shell-like regions at different radii from the central star \cite[e.g.,][]{1990ApJ...360L..51R,2004evn..conf..169A}.

For example, the 43~GHz SiO maser line is frequently detected in a spherical shell region located between the photosphere and the dust formation layer, typically at a distance of about 2--4 times the photospheric radius \cite[e.g., see][]{1994ApJ...430L..61D,2010PASJ...62..431I}. By contrast, the 22.235~GHz H$_2$O maser line is often detected in regions about 5--20 times the photospheric radius \cite[e.g., see][]{1994MNRAS.270..958Y,2011A&A...525A..56R}. In this region, radiation pressure from the central star acts on dust grains, accelerating them outward. The molecular gas is likewise accelerated outward, being entrained by these dust grains.

OH maser lines are typically detected in regions located several hundred times the photospheric radius from the central star \cite[e.g., see][]{2017AJ....153..119O}. Therefore, in spherically symmetric, oxygen-rich CSEs, the 1612~MHz OH maser satellite line (hereafter referred to as the satellite line) is emitted from regions outside those of the 22.235~GHz H$_2$O maser line. By contrast, the 1665/1667~MHz OH maser main lines (hereafter referred to as the main lines) are known to originate from relatively warmer and denser regions closer to the central star than the satellite line \cite[see][]{1989RPPh...52..881C}. In the emission regions of the satellite line, outward acceleration of the gas has ceased and the terminal velocity has been reached; therefore, OH maser lines often exhibit a characteristic double-peaked spectral profile.

By comparing the spectral profiles (especially the velocity widths) of maser lines from different molecular species and transitions, it is possible to identify CSEs with asymmetric kinematic components (AKCs). \citet{2025ApJ...978..114X} referred to cases where maser lines of other molecular species or transitions are detected beyond the velocity range of the satellite line as "velocity excess" and demonstrated that AKC-bearing CSEs can be identified by searching for velocity excess among samples of circumstellar maser sources. In \citet{2025ApJ...978..114X}, the velocity excess of the main line was investigated, and in \citet{2024ApJS..270...13F}, CSEs with AKCs were identified based on the velocity excess of the 22.235~GHz H$_2$O maser line. As noted by \citet{2025ApJ...978..114X}, candidates for circumstellar maser sources with AKCs among CSE samples with effective temperatures below about 300~K include objects of interest for stellar evolution, such as water fountains (WFs), red nova remnants (RNRs), and symbiotic stars. In the present study, we focus in particular on WFs. WFs are evolved, low- to intermediate-mass stars that exhibit collimated, high-velocity bipolar molecular outflows within their CSEs \citep{2007IAUS..242..279I,2012IAUS..287..217D}. The name "water fountain" derives from the observation of high-velocity jets characteristic of WFs via the 22.235~GHz H$_2$O maser line and does not specifically imply an abundance of H$_2$O molecules.

The morphology of the CSE surrounding low- to intermediate-mass evolved stars is thought to change from spherical symmetry to asymmetry in the post-AGB phase. WFs are often assumed to represent a stage marking the onset of this morphological transformation, as pointed out in several studies \cite[e.g.,][]{2008ApJ...689..430S,2012IAUS..287..217D,2015ApJ...799..186G,2019A&A...629A...8T}. The basis for this hypothesis is that the dynamical ages of bipolar molecular outflows observed in WFs are remarkably short, typically on the order of several decades to about 100 years \citep[see][]{2007IAUS..242..279I, 2008ApJ...689..430S}, and that many WFs have been identified as post-AGB stars. More recently, a hypothesis has been proposed that WFs are formed during the common envelope evolution (CEE) process \citep{2022NatAs...6..275K}. In any case, it remains a fact that WF jets are formed during the terminal evolutionary stage of low- to intermediate-mass stars for reasons that remain unclear, and that the formation process is still not well understood. Therefore, a detailed investigation of WFs and their formation mechanisms is valuable for discussing the morphological evolution of CSEs in low- to intermediate-mass stars and the evolution of common envelopes in symbiotic stars.

Traditionally, the identification of WFs has relied on the criterion that the velocity width of the circumstellar 22.235~GHz H$_2$O maser line exceeds 100~km~s$^{-1}$. However, this method may overlook high-velocity WF jets when the line of sight is nearly perpendicular to the jet axis, resulting in a small projection of the velocity along the line of sight. Moreover, WF jets that have just formed may not have accelerated sufficiently, and the velocity width may not reach 100~km~s$^{-1}$, leading to similar oversight. From a purely astronomical perspective, WFs should be defined as dynamically young, high-velocity bipolar molecular jets located at the center of the CSE. If the observational selection criteria are aligned with this astronomical definition, it may be possible to find new WFs that have been overlooked when using only the conventional velocity width (i.e., $\geq 100$~km~s$^{-1}$ of the 22.235~GHz H$_2$O maser line) as the criterion.

As described above, \citet{2024ApJS..270...13F} and \citet{2025ApJ...978..114X} identified CSEs with AKCs through searches for velocity excess. Some of these objects may be WFs that have been missed in previous WF surveys relying solely on the velocity width criterion. Furthermore, among the CSEs with AKCs identified by \citet{2024ApJS..270...13F} and \citet{2025ApJ...978..114X}, there may be WFs with jets that are even dynamically younger than those found by previous methods (see Section~\ref{sec: methodology and Data Origin} for details on the "youth" of WFs).

\citet{2024ApJS..270...13F} compared the velocity widths of the 22.235~GHz H$_2$O maser line and the satellite line using a database of circumstellar H$_2$O and OH masers, identifying 11 CSEs with AKCs. Similarly, \citet{2025ApJ...978..114X} investigated the velocity excess of the main lines and identified 8 CSEs with AKCs. In the present study, we analyze the infrared SEDs of a total of 17 objects (including 2 duplicates) with AKCs identified by \citet{2024ApJS..270...13F} and \citet{2025ApJ...978..114X}, using a one-dimensional dust radiative transfer model, and attempt to find new, young WF candidates embedded within spherically symmetric dust components.

\section{Details of the Methodology and Data Used for Analysis} \label{sec: methodology and Data Origin}

The primary objective of the present study is to identify CSEs with spherically symmetric dust components among the 17 CSEs in which velocity excess (i.e., molecular gas components exhibiting non-spherical motion) has been confirmed by \citet{2024ApJS..270...13F} and \citet{2025ApJ...978..114X}. This work is motivated by the intention to identify WFs at an early stage of their evolutionary sequence. In the present study, we define WFs as either the phenomenon in which a high-velocity bipolar molecular outflow emerges in the central region of a spherically symmetric CSE during the late evolutionary stages of low- and intermediate-mass evolved stars (AGB/post-AGB phase and beyond), or the objects that exhibit this phenomenon. Furthermore, we assume that, in the morphological evolution of the CSEs of low- and intermediate-mass evolved stars, a jet first forms at the center of the spherically symmetric CSE and thereafter the asymmetric changes gradually propagate outward through the entire envelope, driving its evolution. Under this working hypothesis, the presence of spherically symmetric CSEs among circumstellar maser sources showing velocity excess is restricted to the earliest stages of WF evolution (that is, when a small bipolar jet has already formed in the central region, but the outermost layers of the CSE remain unaffected by the jet and preserve a spherical shape). Clarifying whether such objects exist, how many exist if they do, and what properties they exhibit has major implications for understanding the morphological evolution of CSEs during the late evolutionary stages of low- and intermediate-mass stars, and this constitutes the main research question of the present study. In this section, we explore such objects using the methodology described below.

\subsection{Data Used for Analysis and Correction of Interstellar Extinction}\label{sec: data Used for Analysis and Interstellar Extinction Correction}

In this subsection, we describe the preprocessing of the photometric data used in our SED analysis and the correction for interstellar extinction. The photometric data used in the analysis cover a wavelength range from 0.77~$\mu$m to 500~$\mu$m and were obtained from publicly available data archives. The data archives used are as follows: the third generation of the Sloan Digital Sky Survey \cite[SDSS-III,][]{Alam_2015}, Gaia DR2 (the second data release) and EDR3 \citep{2016A&A...595A...1G,2018A&A...616A...1G,2021A&A...649A...1G}, Panoramic Survey Telescope and Rapid Response System \cite[Pan-STARRS DR1,][]{2018AAS...23110201C}, United Kingdom Infra-Red Telescope \cite[UKIDSS DR6,][]{2007MNRAS.379.1599L}, SkyMapper Southern Survey \cite[SkyMapper DR4][]{2018PASA...35...10W}, the second generation of the Palomar Observatory Sky Survey \cite[POSS-II,][]{1963bad..book..481M}, VISTA Near-Infrared YJKs Survey \citep{2006Msngr.126...41E}, Two Micron All Sky Survey the point-source catalog \cite[2MASS PSC,][]{2006AJ....131.1163S}, the bolometric flux estimation for evolved stars \cite[a catalog of 12-band photometric values from 0.55~$\mu$m to 10.2~$\mu$m,][]{2016AJ....152...16V}, Wide-field Infrared Survey Explorer \cite[WISE PSC,][]{2010AJ....140.1868W}, Galactic Legacy Infrared Midplane Survey Extraordinaire \cite[GLIMPSE/IRAC PSC,][]{2003PASP..115..953B}, Midcourse Space Experiment \cite[MSX, version 2.3,][]{2003AAS...203.5708E}, AKARI All-Sky Survey \cite[AKARI PSC][]{2010A&A...514A...1I}, U.S. Air Force Satellite \cite[USAFS,][]{1995yCat.2161....0S}, Infrared Astronomical Satellite \cite[IRAS, version 2.0,][]{1984ApJ...278L...1N,1988iras....1.....B}, and the Herschel InfraRed Galactic Plane Survey \cite[Herschel/PACS PSC,][]{2017MNRAS.471..100E}. Photometric data were collected through the VizieR Catalogue\footnote{\url{https://vizier.cds.unistra.fr/vizier/sed/old/}}.

Regarding the procedure for identifying optical and infrared counterparts, we first cross-matched the positions of maser sources listed in the databases with the WISE Point Source Catalog (PSC) and confirmed, using the WISE colors, that the counterparts were evolved stars with cold CSEs. Our previous studies have confirmed that the positional coordinates given by the WISE PSC are more reliable than those listed in the maser source catalogs \cite[e.g., see details in][]{2024ApJS..270...13F, 2025ApJ...978..114X}. Next, we searched for infrared counterparts at other wavelengths within a radius of 5$''$ centered on the WISE PSC position. Within this search radius, a single counterpart was found for the target object in the bands from 3.35~$\mu$m to 500~$\mu$m. However, in bands shorter than 3.35~$\mu$m, multiple counterparts were sometimes found. Specifically, for IRAS 18251$-$1048, two counterparts were found in Pan-STARRS; for IRAS 18588+0428, two in UKIDSS; for IRAS 19067+0811, two each in Pan-STARRS, SDSS, and UKIDSS; for IRAS 19069+0916, two each in Pan-STARRS and 2MASS; for IRAS 19190+1102, two in UKIDSS; for IRAS 19229+1708, two in Pan-STARRS; and for IRAS 22516+0838, two in Pan-STARRS and five in SDSS were found within the search radius. In such cases, the object closest to the WISE position was adopted as the counterpart for that band. However, it is not possible to completely rule out the possibility that the counterparts identified in these short-wavelength bands are objects other than the target evolved stars. To prevent obvious misidentifications, when plotting the SED, if a data point appeared as a clear outlier relative to adjacent photometric values at neighboring wavelengths, it was regarded as a misidentification and excluded from the analysis. Furthermore, data in catalogs output by VizieR that provide only photometric values without error estimates were excluded from the analysis. Such data represent upper limits and are unsuitable for use in the analysis, regardless of whether the corresponding object is genuine or not. The photometric dataset we compiled is provided as an electronic table accompanying this paper (see Table~\ref{Tab: [SED data]}).

\begin{figure}
		\centering
  		\includegraphics[width=0.5\textwidth, angle=0]{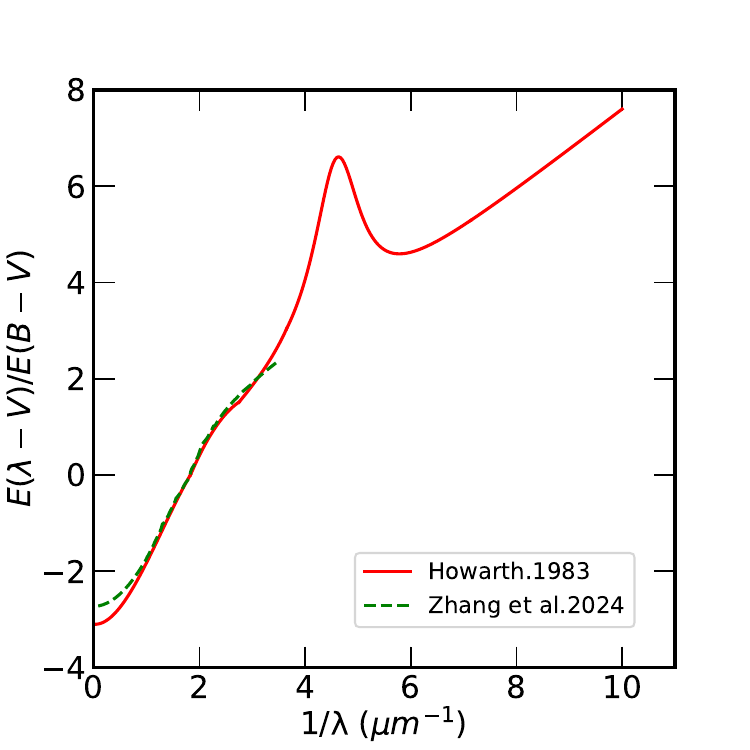}
\caption{Comparison of galactic extinction curves from \citet{1983MNRAS.203..301H} (solid red line) and \citet{2024ApJ...971..127Z} (dashed green line). The curves are plotted as a function of inverse wavelength ($1/\lambda$) on the horizontal axis and normalized color excess $E(\lambda - V)/E(B-V)$ on the vertical axis.}\label{Fig: [the galactic extinction curves]}
\end{figure}

Because interstellar extinction is significant in the short-wavelength range (below 8~$\mu$m), we applied corrections using the Galactic extinction law. The extinction law is defined as:
    \begin{equation} \label{eq-1}
    \frac{E(\lambda-V)}{E(B-V)}\equiv \frac{A_\lambda - A_{\rm V}}{A_{\rm B} - A_{\rm V}}=\frac{A_\lambda - A_{\rm V}}{E(B-V)},
    \end{equation}
where $A_\lambda$, $A_{\rm B}$, and $A_{\rm V}$ represent the extinction at wavelength $\lambda$, the \textit{B}-band, and the \textit{V}-band, respectively, and $E(B-V)$ is the color excess along the line of sight. In the present study, we adopted the infrared extinction law $\frac{E(\lambda-V)}{E(B-V)}$ given by \citet{1983MNRAS.203..301H} (red line in Fig.~\ref{Fig: [the galactic extinction curves]}).

The ratio of total to selective extinction, $R_{\rm V}$, is defined as:
    \begin{equation} \label{eq-2}
        R_{\rm V} \equiv \frac{A_{\rm V}}{A_{\rm B} - A_{\rm V}} = \frac{A_{\rm V}}{E(B-V)}.
    \end{equation}
Here, $A_{\rm V}$ is obtained from the Galactic reddening map by \citet{2011ApJ...737..103S}, available through the NASA/IPAC Infrared Science Archive online tool\footnote{\url{https://irsa.ipac.caltech.edu/applications/DUST/}}. A standard value of $R_{\rm V}=3.1$ was assumed. Therefore, $A_{\lambda}$ can be calculated from equations (1) and (2). $A_{\lambda}$ is defined as:
        \begin{equation} \label{eq-3}
        A_\lambda = -2.5\log_{10}\frac{F_\mathrm{obs}(\lambda)}{F_\mathrm{int}(\lambda)},
    \end{equation}
where $F_\mathrm{obs}$ and $F_\mathrm{int}$ denote the observed and intrinsic fluxes, respectively. Accordingly, the extinction-corrected infrared flux density $F_\mathrm{int}$ can be derived using equation (3).

% In the present study, we confirmed the consistency of the above interstellar extinction values with recent results from \citet{2024ApJ...971..127Z}. Using the latest Gaia XP (Blue Photometer and Red Photometer) slitless spectroscopic data and stellar data from the Large Sky Area Multi-Object Fiber Spectroscopy Telescope \cite[LAMOST,][]{2012RAA....12.1197C,2012RAA....12..723Z}, \citet{2024ApJ...971..127Z} derived Galactic extinction curves over the wavenumber ($1/\lambda$; $\lambda$ in $\mu$m) range 0.01--3.45. Their results are in good overall agreement with those of \citet{1983MNRAS.203..301H} (see Fig.~\ref{Fig: [the galactic extinction curves]}).

In the present study, we confirmed the consistency of the above interstellar extinction values with recent results from \citet{2024ApJ...971..127Z}. Using the latest Gaia XP (Blue Photometer and Red Photometer) slitless spectroscopic data and stellar data from the Large Sky Area Multi-Object Fiber Spectroscopy Telescope \cite[LAMOST,][]{2012RAA....12.1197C,2012RAA....12..723Z}, \citet{2024ApJ...971..127Z} derived Galactic extinction curves over the wavenumber ($1/\lambda$; $\lambda$ in $\mu$m) range 0.01--3.45. Their results are in good overall agreement with those of \citet{1983MNRAS.203..301H} (see Fig.~\ref{Fig: [the galactic extinction curves]}). A small deviation between the two curves can be seen at $1/\lambda < 1$ (i.e., $\lambda \gtrsim 1~\mu$m), corresponding to the near-infrared region, which leads to a slight difference in the extinction correction applied to that wavelength range of the SEDs. However, this discrepancy is minor and does not significantly affect any of the discussions or conclusions presented in this paper.

\subsection{One-Dimensional Dust Radiative Transfer Model}\label{sec: one-dimensional dust radiative transfer model}

To determine whether the outer region of the CSE maintains a spherically symmetric structure, we calculated model SEDs using a one-dimensional (i.e., spherically symmetric) dust radiative transfer model and fitted them to the observed SEDs of individual objects. The purpose of this approach is to judge that, if the model fits well, the dust component of the CSE is likely to be spherically symmetric, whereas a poor fit suggests asymmetry. For this purpose, we employed the publicly available dust radiative transfer code DUSTY \cite[see e.g.,][]{1999astro.ph.10475I,2017MNRAS.465.4482Y}\footnote{\url{https://faculty.washington.edu/ivezic/dusty_web/}}.

The DUSTY model assumes a central point source within a spherically symmetric CSE that emits as a blackbody described by a Planck function at a given temperature. DUSTY incorporates, as default options, optical property data for six representative types of dust grains: “warm” and “cold” silicates \cite[i.e., Sil-Ow, Sil-Oc,][]{1992A&A...261..567O}, silicates and graphite \cite[i.e., Sil-DL, grf-DL,][]{1984ApJ...285...89D}, amorphous carbon \cite[i.e., amC-Hn,][]{1988ioch.rept.....H}, and SiC \cite[SiC-Pg,][]{1988A&A...194..335P}. In models using these standard options for dust grains, the only required input parameter for specifying the dust properties is the relative abundance ratio among the different dust types. Following \citet{2017MNRAS.465.4482Y}, we assumed a 50\%/50\% mixture of “warm” and “cold” silicates \citep{1992A&A...261..567O}. The grain size distribution was assumed to follow the standard Mathis–Rumpl–Nordsieck (MRN) law \cite[MRN:][]{1977ApJ...217..425M}, $n(a)\propto a^{-q}$, with $a_{\rm min}\leq a \leq a_{\rm max}$. The parameter values were set to the standard DUSTY MRN values, namely $q=3.5$, $a_{\rm min}=0.005~\mu$m, and $a_{\rm max}=0.25~\mu$m.

DUSTY provides three options for specifying the radial density distribution: analytic expressions (piecewise power-law or exponential decline), hydrodynamic calculations of radiation-pressure-driven winds (numerical or analytic approximations), and arbitrary user-defined functions. In the present study, considering that the expansion of molecular dust envelopes around AGB stars is driven by radiation pressure on dust grains, we adopted the analytic approximation of radiation-pressure-driven winds. This density distribution offers the advantage of approximating the results of detailed numerical calculations with high accuracy while being computationally efficient. The only required input parameter when using this option is the thickness of the dust shell. We assumed a shell thickness of 10,000 times the inner radius.

The remaining three free parameters are the effective temperature of the central radiation source ($T_{\rm eff}$), the dust temperature at the inner boundary of the dust shell ($T_{\rm d}$; theoretically, if multiple grain species are mixed, the temperatures may differ for each, but DUSTY treats mixtures as a single species with average properties, so only one temperature is specified), and the optical depth at 2.2~$\mu$m ($\tau_{\rm 2.2}$). Thus, by varying these three parameters (i.e., exploring the parameter space), we attempt to fit the DUSTY model to the SEDs of the target objects. The quantitative evaluation method for the fitting and the results are described in Section~\ref{sec: quantitativeassessment of model match and its outcomes}.

When computing numerous models to uniformly cover the parameter space defined by these three parameters, we find that two principal types of model SED profiles are reproduced: single-peaked (S) and double-peaked (D). Single-peaked SED profiles are typically seen when the CSE has not yet detached from the central star and are considered characteristic of typical AGB stars. In contrast, double-peaked SEDs indicate that the envelope has detached from the central star and are interpreted as corresponding to the post-AGB stage. Of the two peaks, the shorter-wavelength (near-infrared) peak is primarily attributed to the stellar photosphere, whereas the longer-wavelength (mid- to far-infrared) peak is interpreted as arising from the isothermal dust component of the detached envelope \cite[e.g.,][]{1989ApJ...346..265H, 1993ARA&A..31...63K}. However, this interpretation reflects only a general tendency; for example, even in more evolved PNe such as IRAS 17347--3139, IRAS 17393--2727, and IRAS 19255+2123, single-peaked SEDs can still be observed \citep{2012A&A...547A..40U}.

The essence of our analysis is that, if the SED of a target object is well fitted by the DUSTY model, we judge the CSE to be spherically symmetric. However, it is also possible for a nearly spherical but slightly asymmetric CSE to be fitted well by the DUSTY model. For instance, the object IRAS 22272+5435 is known to contain a geometrically thick torus in its CSE \citep{2012ApJ...759...61N}, but \citet{2017MNRAS.465.4482Y} demonstrated that the observed SED can nevertheless be fitted satisfactorily by the DUSTY model (spherical symmetry). Conversely, it is difficult to envisage a case in which a model fails to fit well while the CSE is actually spherical. Therefore, fitting with a one-dimensional model can be regarded as a reasonably effective probe for investigating the morphology of CSEs.

It should be noted that, aside from geometric assumptions, some physical factors may also influence the results of SED fitting. The primary contributors would be spectral lines and bands arising from dust and molecular components. Nevertheless, these effects are expected to have only a minor impact on the present analysis. First, the observed SEDs used in the present study are constructed from photometric data in broadband filters, in which local line and band features are averaged out. Second, the standard optical constants of dust adopted in the present study have already been shown to reproduce the SED shapes of AGB and post-AGB stars \citep[e.g.,][]{1992A&A...261..567O,1984ApJ...285...89D,2017MNRAS.465.4482Y}. Therefore, although the influence of dust and molecular components on the SED cannot be completely excluded, within the scope of the present study their effect on the one-dimensional DUSTY fitting results can be regarded as negligible.

% 修正済みLaTeX原稿をここに記載
\subsection{Quantitative Assessment of Model Fit and Its Outcomes}\label{sec: quantitativeassessment of model match and its outcomes}

As a quantitative criterion for assessing whether the DUSTY model reproduces the observed SED well, in the present study we adopted the reduced $\chi^2$ value used by \citet{2025MNRAS.539.1220D} (hereafter simply referred to as the $\chi^2$ value). The $\chi^2$ value in this analysis was calculated as:
    \begin{equation} \label{eq-4}
        \frac{\chi^2}{\rm dof}= \frac{\sum_{i}(O_i - E_i)^2/E_i^2}{N_{\rm obs} - N_{\rm para}},
    \end{equation}
Here, $O_{i}$ denotes the observed value, and $E_{i}$ denotes the value predicted by the DUSTY code. The abbreviation dof stands for degrees of freedom, $N_{obs}$ is the number of observational data points, and $N_{ para}$ is the number of model parameters.

Using DUSTY, we computed model SEDs over the parameter space shown in Table~\ref{Tab: [The_models_generated_by_DUSTY_codes_in_different_parameters]}, and calculated the $\chi^2$ values between the resulting model SEDs and the observed SEDs. Subsequently, we selected the model within the parameter space that yielded the minimum $\chi^2$ value, and then further minimized the $\chi^2$ value by shifting the model SED vertically (i.e., along the flux axis). The $\chi^2$ value obtained in this way was adopted as the final criterion for evaluating the goodness of fit. In addition, to verify the consistency between the observed and model SEDs, we carried out visual inspections and regarded the fit as acceptable when the $\chi^2$ value was less than $9.00 \times 10^{-4}$. Ultimately, for seven targets, we obtained model SEDs with $\chi^2$ values below $9.00 \times 10^{-4}$. The results of the model fits are summarized in Table~\ref{Tab: [list of objects used in this study and Dusty parameters]}. These seven objects are candidates in which the outermost CSE is spherically symmetric, whereas asymmetric molecular gas motions may be present in the central regions.

\section{Results}\label{sec: results}

\begin{figure}
		\centering
  		\includegraphics[width=0.8\textwidth, angle=0]{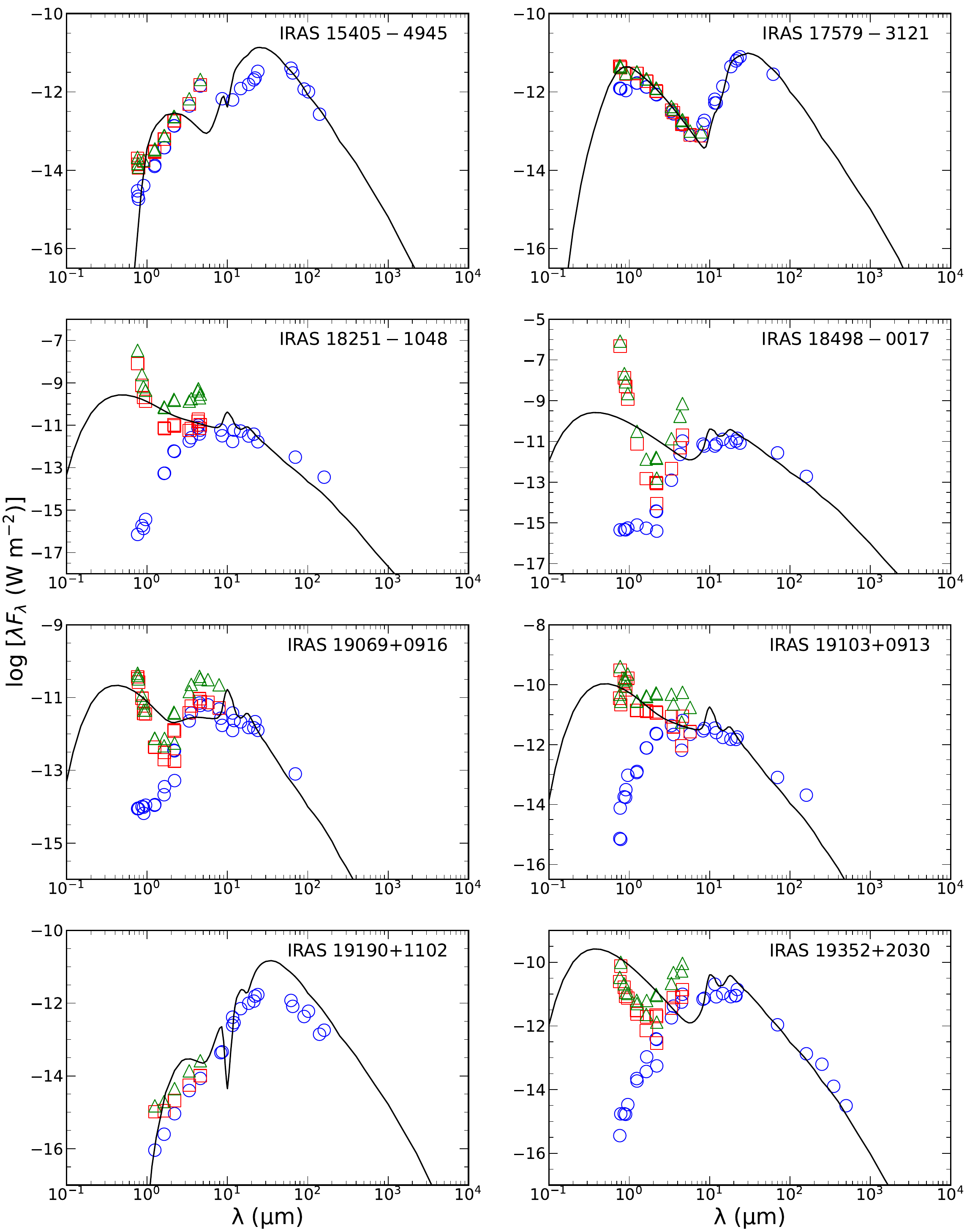}
\caption{
Spectral energy distributions (SEDs) of objects with velocity excess in the OH mainlines, based on \citet{2025ApJ...978..114X}.
The blue circles show the observed data points without interstellar extinction correction.
The red squares indicate the data points corrected for extinction using the law of \citet{1983MNRAS.203..301H} (applied only to wavelengths shorter than 8~$\mu$m), while the green triangles represent the corrections using the law of \citet{2024ApJ...971..127Z} (see Section~\ref{sec: data Used for Analysis and Interstellar Extinction Correction} for details).
The black lines represent the best-fit models obtained with the DUSTY code (see text for details).
The complete dataset is provided in the electronic table accompanying this paper.
}\label{Fig: [spectral energy distributions of objects with velocity excess in the OH 1665/1667 MHz lines]}
\end{figure}

\begin{figure}
		\centering
  		\includegraphics[width=0.8\textwidth, angle=0]{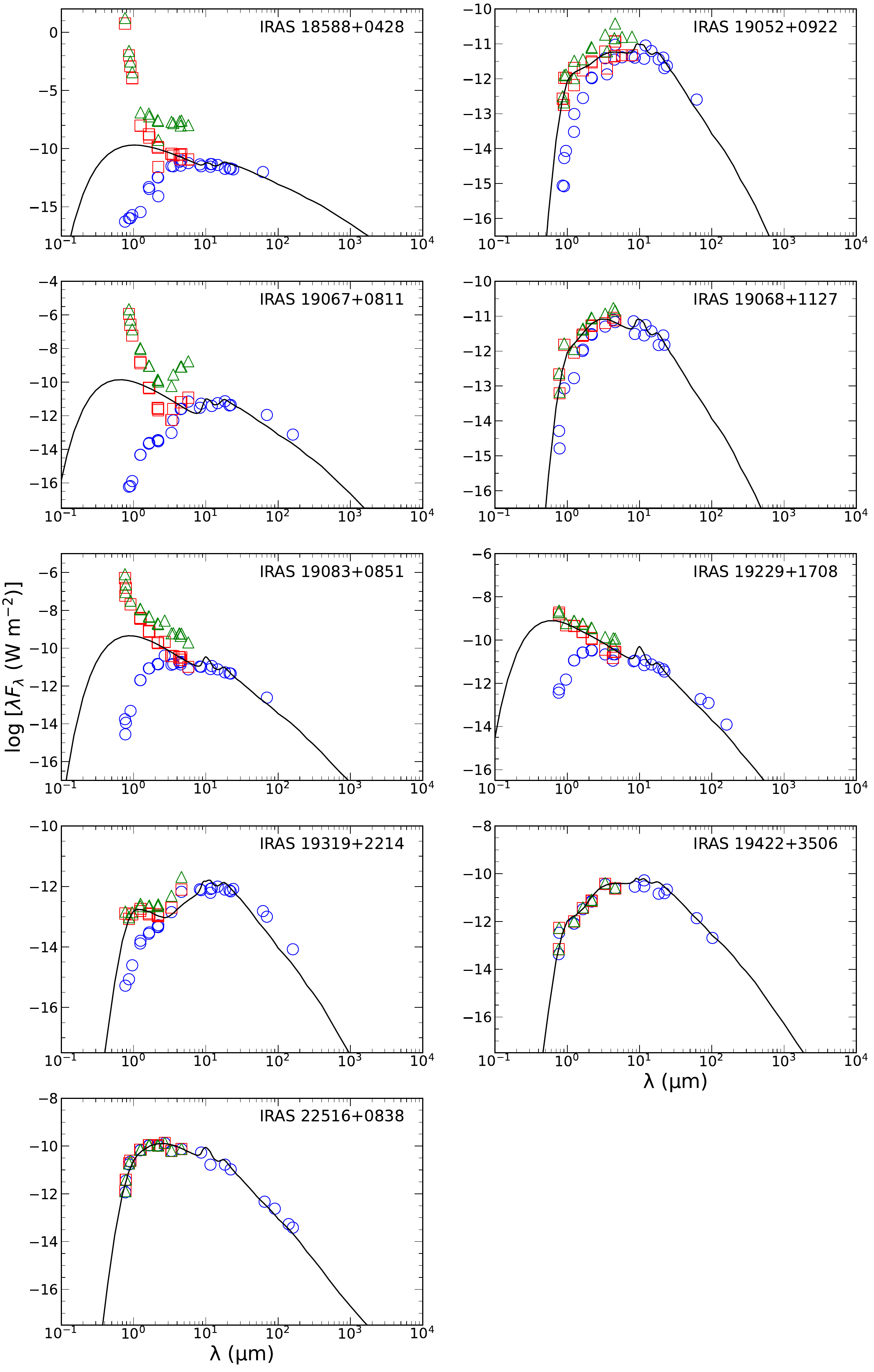}
\caption{
Spectral energy distributions (SEDs) of Water Fountain candidates from \citet{2024ApJS..270...13F}. The notations for data points and model fits are the same as those in Figure~\ref{Fig: [spectral energy distributions of objects with velocity excess in the OH 1665/1667 MHz lines]}; see the caption of Figure~\ref{Fig: [spectral energy distributions of objects with velocity excess in the OH 1665/1667 MHz lines]} for details.
}
\label{Fig: [spectral energy distributions of Water Fountain candidates]}
\end{figure}

Among the 17 CSEs with AKCs (11 from \citet{2024ApJS..270...13F} and 8 from \citet{2025ApJ...978..114X}, with two overlapping objects, namely IRAS 18251$-$1048 and IRAS 19069+0916), modeling with DUSTY yields good fits for seven cases (see Figures~\ref{Fig: [spectral energy distributions of objects with velocity excess in the OH 1665/1667 MHz lines]} and \ref{Fig: [spectral energy distributions of Water Fountain candidates]}, where the best-fitting SED models are presented). Columns 5-–11 of Table~\ref{Tab: [list of objects used in this study and Dusty parameters]} present their SED-fitting results, including the SED profile type, effective temperature ($T_{\rm eff}$), dust temperature ($T_{\rm d}$), optical depth at 2.2~$\mu$m ($\tau_{\rm 2.2}$), the mass-loss rate ($\dot{M}$), the upper limit to the stellar mass ($M_{\rm star}$), and the $\chi^2$ values, respectively. The lack of corresponding information (i.e., cases that do not yield a good fit with DUSTY) is indicated by an ellipsis (\dots) in the respective columns. 
Among the seven cases that were well fitted, two objects—IRAS 17579$-$3121 and IRAS 19319+2214—exhibit double-peaked SED profiles, while the remaining five objects (IRAS 19052+0922, IRAS 19068+1127, IRAS 19229+1708, IRAS 19422+3506, and IRAS 22516+0838) display single-peaked SED profiles. In general, a single-peaked SED indicates that the CSE has not yet detached from the stellar photosphere, which is a characteristic feature commonly observed in AGB stars. In contrast, a double-peaked SED suggests that the envelope has separated from the photosphere of the central star, marking the post-AGB phase. The near-infrared peak of a double-peaked SED mainly originates from the reddened photosphere of the central star, whereas the mid-infrared peak is produced by cool dust within the detached envelope \cite[e.g.,][]{1989ApJ...346..265H, 1993ARA&A..31...63K}. Conversely, the remaining ten objects did not yield satisfactory SED fits with the DUSTY model.

\subsection{Objects with Single-peaked SEDs}\label{sec: single-peaked SED profiles}

Five WF candidates from \citet{2024ApJS..270...13F}—IRAS 19052+0922, IRAS 19068+1127, IRAS 19229+1708, IRAS 19422+3506, and IRAS 22516+0838—exhibit single-peaked SED profiles that are well fitted by DUSTY models. The DUSTY model successfully reproduces the overall spectral shape and the characteristic single-peaked profile. Therefore, we consider these cases to be well fitted. The best-fitting model parameters show effective temperatures from 2000 to 10000~K and dust temperatures between 850 and 1250~K. Among them, IRAS 19422+3506 exhibits a peak in the mid-infrared (at $\sim$12~$\mu$m). By contrast, the other four sources peak at shorter wavelengths: IRAS 19229+1708 and IRAS 22516+0838 at approximately 2~$\mu$m, while IRAS 19052+0922 and IRAS 19068+1127 peak near 4.6~$\mu$m. IRAS 19229+1708 and IRAS 19422+3506 exhibit relatively large optical depths, indicative of thicker circumstellar dust shells. The presence of relatively hot dust ($T_{\rm d}=850$--$1250$~K) in these objects suggests that the envelope remains closely associated with the central star.

\subsection{Objects with Double-peaked SEDs}\label{sec: double-peaked SED profiles}

Two objects, IRAS 17579$-$3121 and IRAS 19319+2214, exhibit double-peaked SED profiles that can be well fitted using the radiative transfer code DUSTY. For IRAS 17579$-$3121, its double-peaked SED, with clear peaks at approximately 1~$\mu$m and 30~$\mu$m, is well reproduced by DUSTY models. The well-separated infrared peaks are indicative of substantial dusty circumstellar material. The derived dust temperature of the envelope, 100~K, falls within the typical range for single post-AGB stars \cite[see][]{1993ARA&A..31...63K}. Our derived $T_{\rm eff}$ (6600~K) is roughly consistent with 6500~K reported by \citet{2014RMxAA..50..293M}. These temperatures are reminiscent of the very young post-AGB star IRAS 22272+5435 with $T_{\rm eff}=8200$~K and $T_{\rm d}=200$~K \cite[which left the AGB about 380~years ago; see][]{2001ApJ...557..831U}, whose double-peaked SED profile is also well fitted by DUSTY models \citep{2017MNRAS.465.4482Y}. It also exhibits a relatively high mass-loss rate of $1.08\times10^{-4}$~M$_{\odot}$~yr$^{-1}$, further supporting the presence of a dense dusty envelope.

Another object, IRAS 19319+2214, exhibits a double-peaked SED profile with peaks around 1~$\mu$m and 15~$\mu$m and is also well fitted by DUSTY models. Compared to IRAS 17579$-$3121, IRAS 19319+2214 exhibits a higher dust temperature of 500~K (see Table~\ref{Tab: [list of objects used in this study and Dusty parameters]}). The presence of hot dust in post-AGB stars may indicate binarity \citep{2003ARA&A..41..391V}.

\subsection{Objects with Unreliable SED Fits}\label{sec: SED profiles with Pporly constrained Fits}

Eight objects—IRAS 15405$-$4945, IRAS 18251$-$1048, IRAS 18498$-$0017, IRAS 18588+0428, IRAS 19067+0811, IRAS 19069+0916, IRAS 19083+0851, and IRAS 19103+0913—exhibit poor SED fits across the entire wavelength range. Except for IRAS 15405$-$4945, the near-infrared data for these seven sources may be significantly affected by interstellar extinction, contributing to the poor fits. Among them, IRAS 15405$-$4945, IRAS 18498$-$0017, IRAS 19067+0811, and IRAS 19069+0916 display double-peaked SED profiles. Moreover, in the sources IRAS 18498$-$0017, IRAS 19067+0811, and IRAS 19069+0916, there is a hint that the emission flattens or rises again at $\lesssim 2~\mu\mathrm{m}$ (based on the original data). Owing to the strong extinction, even the extinction-corrected near-infrared features may not be trustworthy. Additionally, a distinct trend emerges wherein the mid- to far-infrared portions of the SED can be reasonably well fitted by DUSTY, while the near-infrared data remain inconsistent with the model. Two objects—IRAS 19190+1102 (WF) and IRAS 19352+2030 (with substantial interstellar extinction in the near-infrared)—fall into this category. These cases are also classified as poorly fitted objects.

\section{Discussion}
\label{sec: discussion}

\subsection{Distribution in the IRAS Two-Color Diagram}
\label{sec: overall distribution in the IRAS colour-colour diagram}

\begin{figure}
		\centering
  		\includegraphics[width=0.8\textwidth, angle=0]{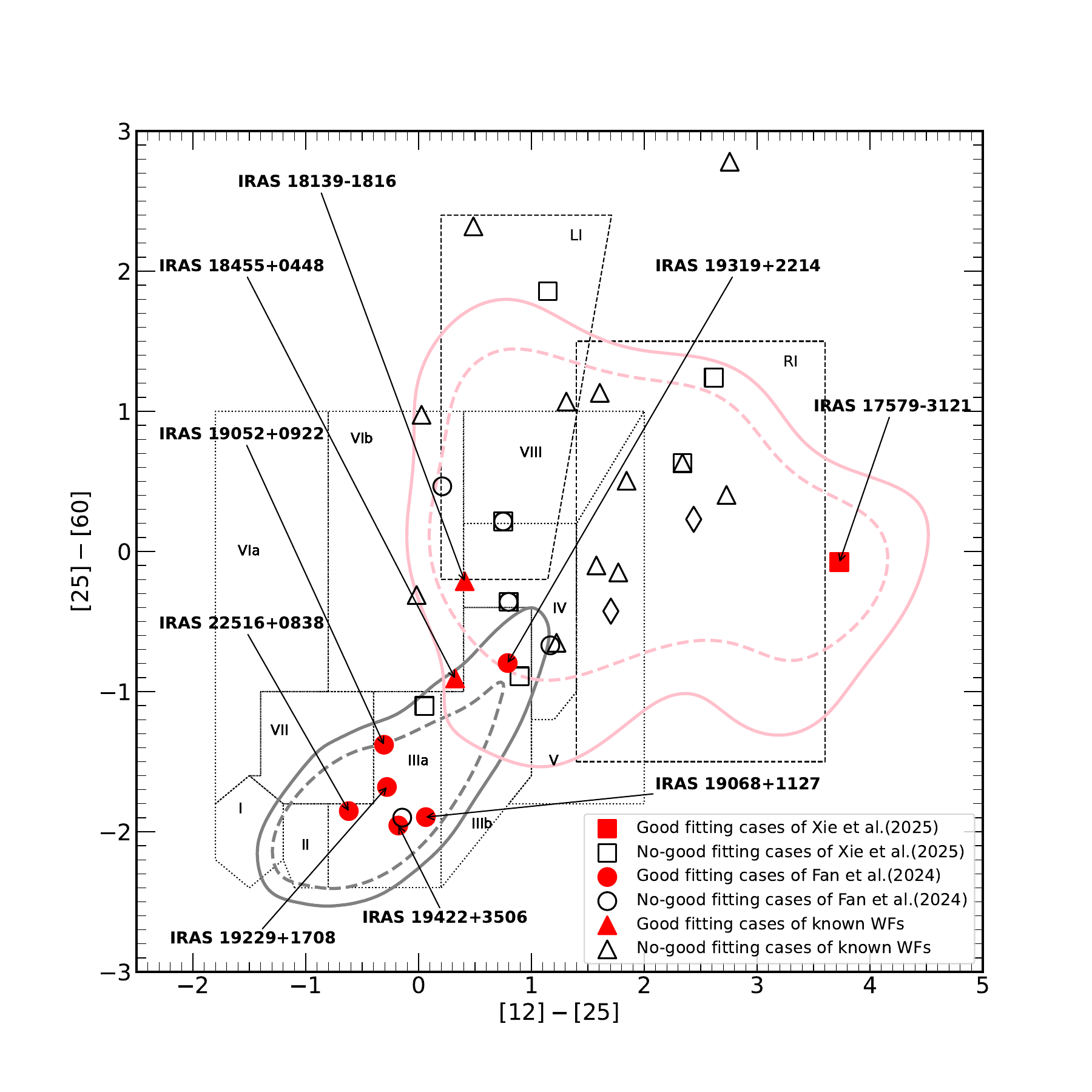}
\caption{IRAS two-color diagram using [12]$-$[25] versus [25]$-$[60] displays the good (red filled marks) and poor (open marks) DUSTY fitting cases from \citet{2025ApJ...978..114X}, \citet{2024ApJS..270...13F}, and \citet{2017MNRAS.465.4482Y}, respectively. The colour indexes are defined as [12]$-$[25]=$2.5\log(F_{\text{25$\mu m$}}/F_{\text{12$\mu m$}})$ and [25]$-$[60]=$2.5\log(F_{\text{60$\mu m$}}/F_{\text{25$\mu m$}})$. The boxes (black dotted lines) are defined by \citet{1988A&A...194..125V}. The O-rich AGB (grey lines) and post-AGB (pink lines) regions are defined by \citet{2025ApJ...978..114X}. The regions LI and RI (black dashed lines) represent two post-AGB sequences \citep{2002AJ....123.2772S,2002AJ....123.2788S}. Red filled and open squares represent good and poor DUSTY fitting cases from \citet{2025ApJ...978..114X}, respectively, while red filled and open circles indicate the same from \citet{2024ApJS..270...13F}. Red filled triangles mark well-fitted known WF sources, and open triangles indicate poorly-fitted ones. The open diamonds indicate two WFs (IRAS 15103$-$5754 and IRAS 17291$-$2147) not included in the results of \citet{2017MNRAS.465.4482Y} (see the texts). }\label{Fig: [IRAS two-color diagram of good and no-good DUSTY fitting cases]}
\end{figure}

Figure~\ref{Fig: [IRAS two-color diagram of good and no-good DUSTY fitting cases]} shows the locations of the objects analyzed in the present study and known WF sources in the IRAS two-color diagram. They are classified by the quality with which a one-dimensional DUSTY model reproduces their SEDs. Among the 16 known WFs, SED fitting results for 14 sources have already been reported by \citet{2017MNRAS.465.4482Y}. In that study, only two objects, IRAS 18139$-$1816 and IRAS 18455+0448, were reported to be well reproduced by a single-peaked one-dimensional DUSTY model. For the two WFs not included in \citet{2017MNRAS.465.4482Y}, we performed new SED fitting and confirmed that both could not be reproduced by a one-dimensional model (the fitting results are presented in Appendix~\ref{appendix: two WFs not included in the results of Yung et al. (2017)}).

In the IRAS two-color diagram, regions I-–VIII enclosed by black dashed lines correspond to various types of evolved stars at different evolutionary stages. Regions I–-IV are mainly occupied by oxygen-rich stars, region V by planetary nebulae (PNe), regions VI and VII by carbon-rich stars, and regions IV, V, and VIII primarily by post-AGB stars \cite[e.g., see][]{1988A&A...194..125V,1998A&AS..127..185N,2002A&A...388..252E}. Furthermore, the regions labeled LI and RI \citep{2002AJ....123.2772S,2002AJ....123.2788S} represent post-AGB stars with different expansion velocities and initial masses. LI sources tend to have larger expansion velocities and higher initial masses than RI sources and are thought to have left the AGB phase at an earlier stage. These LI sources are inferred to evolve into bipolar planetary nebulae after experiencing irregular mass loss during the late post-AGB phase, whereas low-mass RI sources are believed to evolve into elliptical planetary nebulae \citep{2002AJ....123.2772S}. The gray contours indicate the boundaries of the AGB region, and the red contours represent those of the post-AGB region \cite[for details, see][]{2025ApJ...978..114X}.

Overall, objects with good fits are located mainly in regions showing AGB-like colors, whereas those with poor fits are distributed in regions showing post-AGB-like colors. This finding is consistent with our initial expectation. As mentioned earlier, most of the objects with good fits are expected to have spherically symmetric CSEs, a characteristic feature of AGB stars. In contrast, objects with poor fits are considered to possess non-spherical structures (e.g., bipolar jets or tori) within their CSEs—features typically observed in more evolved post-AGB stars and PNe.

According to \citet{2007ApJ...663..342H}, jets appear within several hundred years after torus formation, with an average delay of about 300 years. Observationally, a torus is found to form first, followed by the development of jets, indicating a clear sequence of morphological evolution. Thus, at the evolutionary stage when jets begin to form, a torus is already present in the CSE, and both structures coexist. If the torus is geometrically thick, its shape is relatively close to spherical, and at an early stage—before it becomes large—the deviation from spherical symmetry may be small. Similarly, if the jets are newly formed and very small in size, their contribution to the SED is expected to be minor. In such cases, even when a torus or jets exist within the CSE, it may still be possible to obtain a good fit using a one-dimensional model. For example, IRAS 22272+5435 is interpreted as having a spherically expanding envelope and a torus \citep[][no jet has been detected]{2012ApJ...759...61N}, yet its SED is known to be well fitted by a one-dimensional DUSTY model \citep{2017MNRAS.465.4482Y}.

The two known WFs that are well fitted by DUSTY (IRAS 18139$-$1816 and IRAS 18455+0448), as well as IRAS 19319+2214, have observationally confirmed H$_2$O maser velocity excesses, indicating that asymmetric structures (tori or jets) are almost certainly present near the central star within the CSE. Nevertheless, their infrared SEDs can still be reproduced by a one-dimensional DUSTY model. This fact suggests that these objects represent systems in which a torus and nascent jets coexist at the center of the CSE—an early phase of morphological evolution as proposed by \citet{2007ApJ...663..342H}. From this viewpoint, the results shown in Figure~\ref{Fig: [IRAS two-color diagram of good and no-good DUSTY fitting cases]} are consistent with that evolutionary scenario.

However, IRAS 17579$-$3121 represents a somewhat exceptional case. Despite showing extremely red infrared colors (outside the post-AGB region), its SED is well reproduced by a one-dimensional model. We briefly discuss possible reasons for this behavior below. As illustrated by IRAS 22272+5435, even a CSE with asymmetric structures such as tori or jets can sometimes be reproduced by a one-dimensional DUSTY model. The success of such fitting likely depends on the shape and size of the torus and the viewing angle. Indeed, IRAS 07134+1005, which has a CSE structure similar to that of IRAS 22272+5435, cannot be well reproduced by a one-dimensional DUSTY model \citep{2017MNRAS.465.4482Y}. One of the key differences between IRAS 22272+5435 and IRAS 07134+1005 lies in the torus size and dynamical timescale. The torus of IRAS 22272+5435 is several times smaller than that of IRAS 07134+1005 \citep{2009ApJ...692..402N,2012ApJ...759...61N}. The dynamical timescales of the inner and outer edges of the torus in IRAS 22272+5435 are estimated to be about 420 and 1100 years, respectively \citep{2012ApJ...759...61N}, whereas those in IRAS 07134+1005 are estimated to be 1140–-1710 years (inner edge) and approximately 4270 years (outer edge) \citep{2009ApJ...692..402N}. Thus, in IRAS 07134+1005, where the torus is older and larger, the torus contribution to the total infrared emission is more significant, preventing a good fit with a simple spherical model.

The DUSTY modeling results for IRAS 17579$-$3121 indicate a high mass-loss rate of $1.08\times10^{-4}$~M$_\odot$~yr$^{-1}$, consistent with a superwind mass-loss rate \cite[e.g., see][]{1992ApJ...397..552W}. Such a high mass-loss rate may produce infrared colors similar to those of PNe, while the superwind phenomenon itself is characteristic of the terminal AGB phase \citep{1993ApJ...413..641V}. Superwinds are generally non-spherical, being stronger in the equatorial plane than in the polar direction; this equatorially enhanced mass loss is considered the primary cause of torus formation \cite[e.g.,][]{1996A&A...313..605D,2000MNRAS.312..217S}. Therefore, assuming that IRAS 17579$-$3121 is a young post-AGB star in the superwind phase (with a still-small torus) observed at a nearly edge-on angle, both its very red color and its successful one-dimensional fit can be naturally explained.

Another point worth noting for IRAS 17579$-$3121 is the presence of another nearby IRAS source, IRAS~17580$-$3111. \citet{2008AJ....135.2074G} detected an OH maser line in the direction of IRAS~17579$-$3121, but its position coincides, within the measurement uncertainties, with that of IRAS~17580$-$3111 (R.A.(J2000) = 18$^{\mathrm{h}}$01$^{\mathrm{m}}$20$^{\mathrm{s}}$.4, Dec.(J2000) = $-$31$^\circ$11$'$20$''$.3). Thus, it is uncertain which IRAS source actually produced the detected maser emission. In fact, IRAS~17580$-$3111 has IRAS colors of [12]$-$[25]=1.7 and [25]$-$[60]=$-$0.7, typical of post-AGB stars.

\subsection{Exploring Evolutionary Stages via Infrared Variability}
\label{sec: young WF candidates}

To clarify the evolutionary stages of our sample, we analyzed the IRAS variability indices and WISE infrared light curves. A high IRAS variability index and a periodic variation in the WISE bands strongly support an AGB classification. Period analyses were performed for all objects shown in Figure~\ref{Fig: [IRAS two-color diagram of good and no-good DUSTY fitting cases]}. For the sample analyzed by \citet{2025ApJ...978..114X}, period analyses had already been conducted. In the present study, we performed new analyses for the objects listed in \citet{2024ApJS..270...13F} and known WF sources.

\begin{figure}
\centering
\includegraphics[width=\textwidth, angle=0]{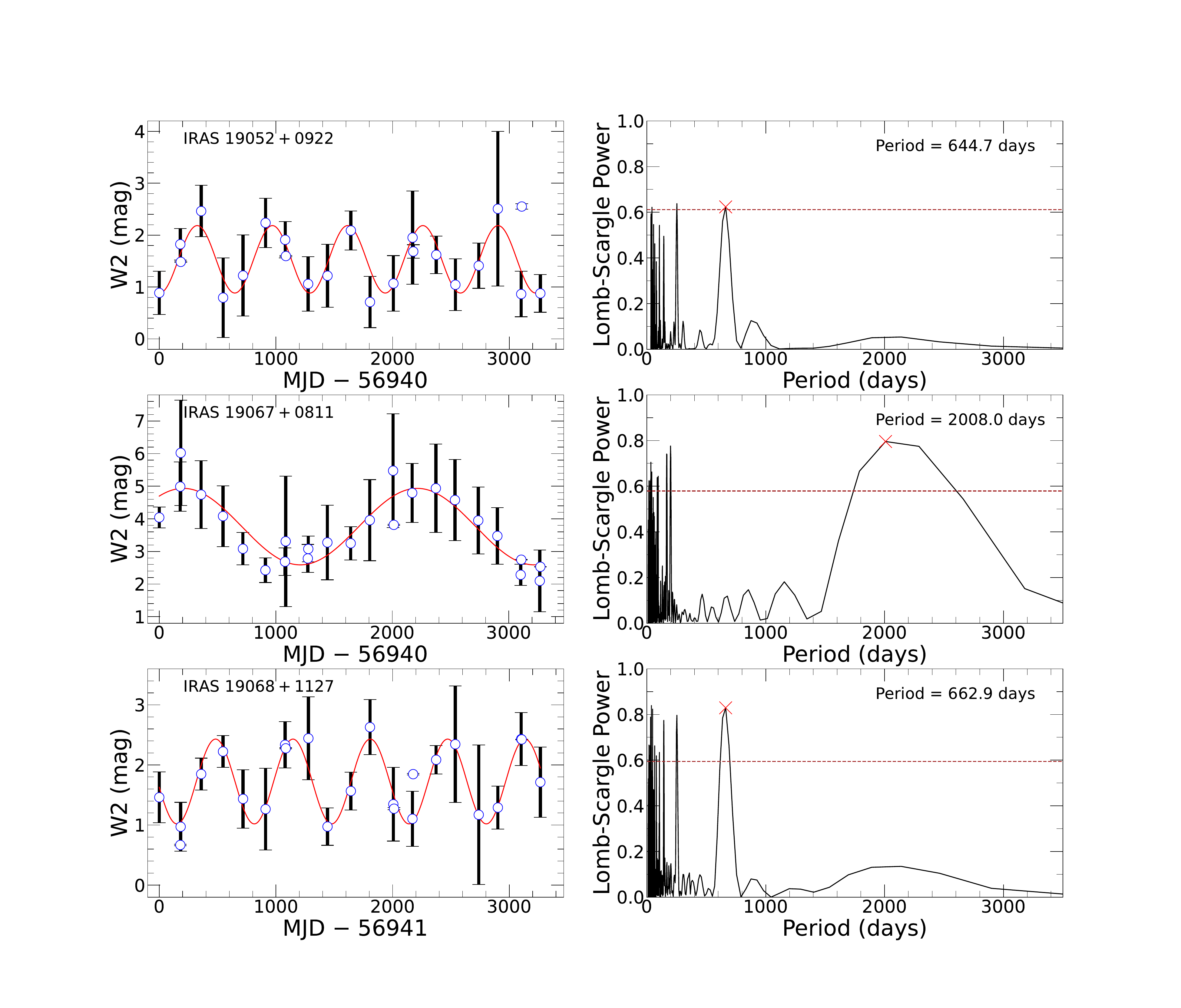}
\caption{WISE light curves (left panel) and periodic analysis results (right panel) for the sources of \citet{2024ApJS..270...13F} for which periodicity was detected. The red curve in each WISE light curve represents the best-fit sinusoid based on the Lomb–Scargle periodogram. The corresponding fit results are: 0.65sin(2$\pi$(ft-0.25))+1.53, 1.17sin(2$\pi$(ft+0.15))+3.76, and --0.71sin(2$\pi$(ft+0.02))+1.72, where $f$ is the frequency equal to 1/period. The black vertical lines represent the error bars. In the Lomb–Scargle periodogram, the dashed brown horizontal line indicates the periodogram level corresponding to a maximum peak false alarm probability of 2\%, and the red X mark denotes the selected peak.}
\label{Fig: [WISE light curves with periodicity for the sources of Fan et al.2024]}
\end{figure}

\begin{figure}
		\centering
  		\includegraphics[width=\textwidth, angle=0]{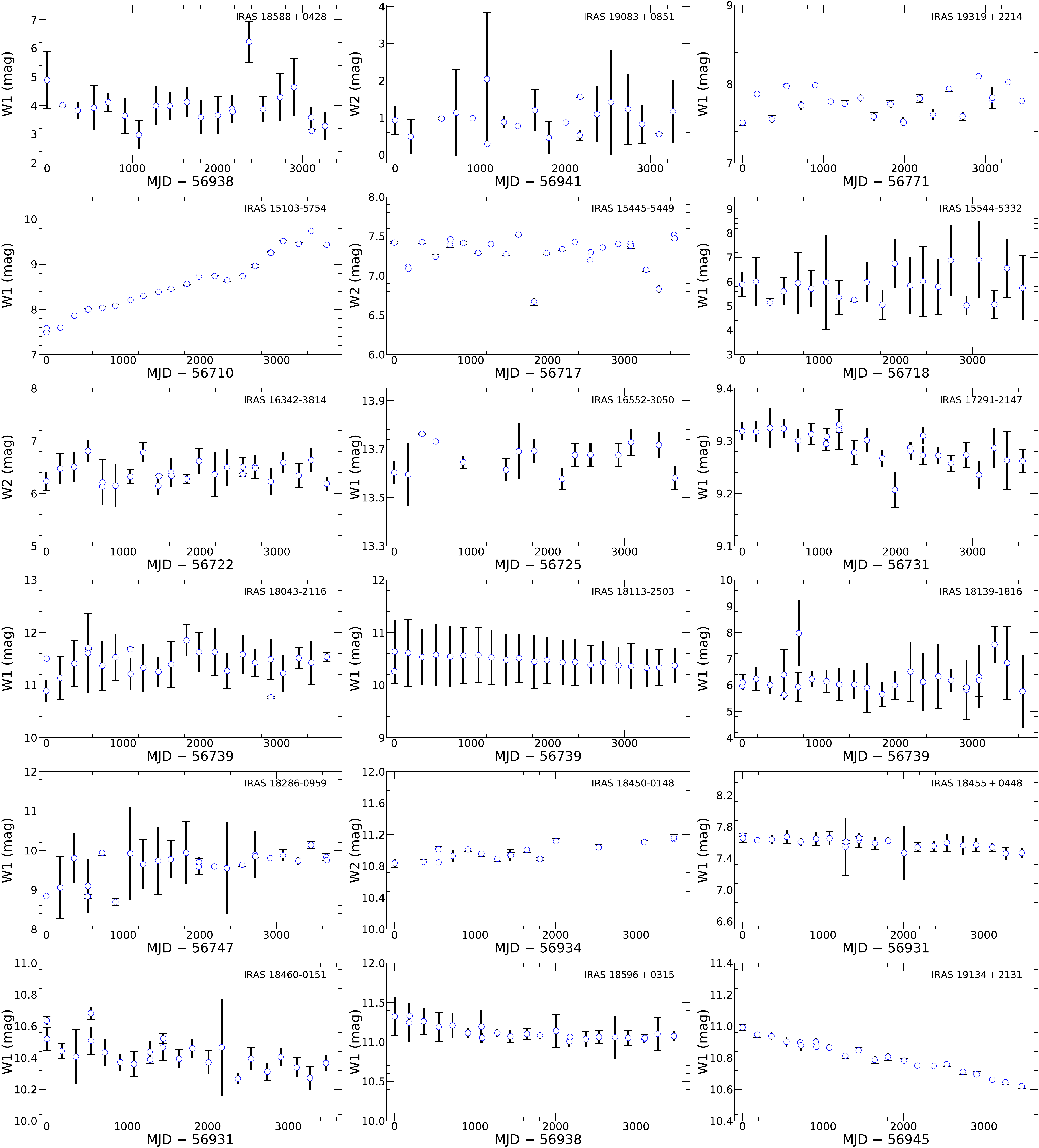}
    	\caption{WISE light curves of objects with no detectable periodicity (see the main text for details). The object names are shown in the upper right corner; MJD denotes the modified Julian day, and the W1 and W2 bands are plotted together.}
	\label{Fig: [WISE light curves of known WFs]}
\end{figure}

\begin{figure}
		\centering
  		\includegraphics[width=0.8\textwidth, angle=0]{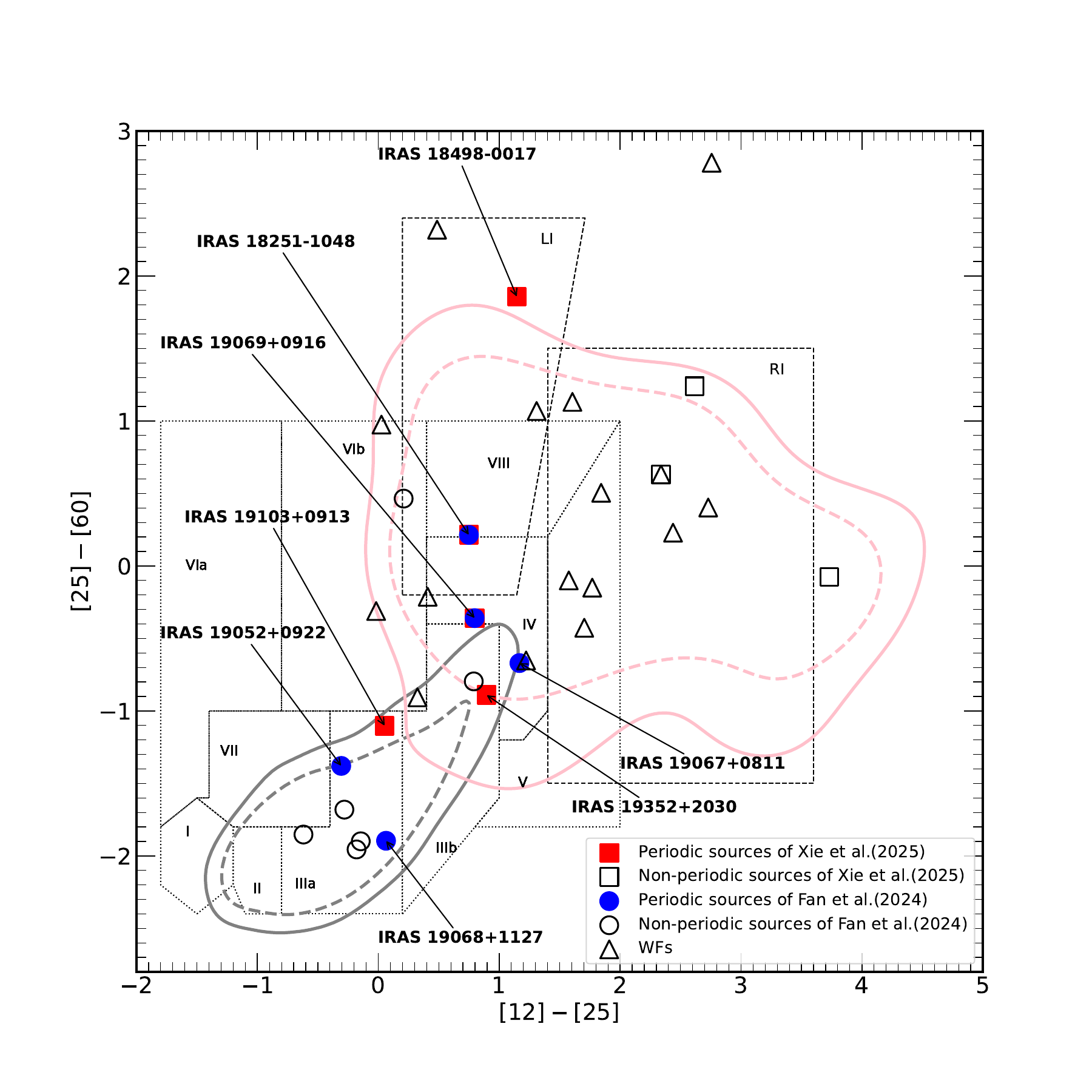}
\caption{Distributions of the periodic and non-periodic sources from our sample and known WFs in the IRAS two-color diagram. Red filled and open squares represent the periodic and non-periodic sources from \citet{2025ApJ...978..114X}, respectively, while blue filled and open circles indicate the same from \citet{2024ApJS..270...13F}. The open triangles mark the non-periodic WFs (see the texts). The boxes are the same as in Figure~\ref{Fig: [IRAS two-color diagram of good and no-good DUSTY fitting cases]}.}\label{Fig: [IRAS two-color diagram of periodic and non-periodic sources]}
\end{figure}

The IRAS variability indices for our targets are given in Column~2 of Table~\ref{Tab: [IRAS variability index and WISE light curve for our target objects]}. Larger values indicate a higher probability that the object is variable. To investigate possible periodicity, we constructed WISE light curves using high-quality W1 and W2 photometric data from the NEOWISE-R mission and applied Lomb–Scargle periodogram analyses \cite[for details, see][]{2025ApJ...978..114X}. For cases where periodicity was found, the corresponding WISE light curves are shown in Figure~\ref{Fig: [WISE light curves with periodicity for the sources of Fan et al.2024]}. The pulsation periods of IRAS 19052+0922, IRAS 19067+0811, and IRAS 19068+1127 were derived to be 644.7, 2008.0, and 662.9~days, respectively. These results are summarized in Column~4 of Table~\ref{Tab: [IRAS variability index and WISE light curve for our target objects]}. On the other hand, WISE light curves of objects without statistically significant periodicity are collectively shown in Figure~\ref{Fig: [WISE light curves of known WFs]}. Period analyses of known WFs also revealed no clear periodicity. The distributions of periodic and non-periodic sources in the IRAS two-color diagram are shown in Figure~\ref{Fig: [IRAS two-color diagram of periodic and non-periodic sources]}. In general, periodicity tends to disappear as evolution proceeds, although the presence or absence of periodicity does not always correspond to the color boundary distinguishing AGB and post-AGB stars (i.e., the boundary between regions IIIa and IIIb).

From the perspective of morphological evolution discussed in Section~\ref{sec: overall distribution in the IRAS colour-colour diagram}, objects whose SEDs are well reproduced by one-dimensional DUSTY models (i.e., those with good fits) can be interpreted as being younger than those with poor fits. Based on their infrared colors and variability characteristics, well-fitted objects can be subdivided into three groups: (1) pulsating objects with AGB-like colors (IRAS 19052+0922 and IRAS 19068+1127); (2) non-pulsating objects with AGB-like colors (IRAS 19229+1708, IRAS 19422+3506, and IRAS 22516+0838); and (3) non-pulsating objects with post-AGB-like colors (IRAS 17579$-$3121 and IRAS 19319+2214). All known WFs show no pulsations and exhibit post-AGB-like (or PN-like) infrared colors, corresponding to category (3). Objects in categories (1) and (2) may thus represent earlier evolutionary stages than known WFs, based on their variability and infrared characteristics.

Objects in category (2), which show no pulsation but exhibit AGB-like infrared colors, can be interpreted as retaining thick circumstellar dust shells typical of AGB stars, while the stellar pulsation and mass loss have already weakened or ceased. This corresponds to a very short transitional phase immediately after the termination of mass loss at the end of the AGB phase. However, interpretation based on infrared color alone always involves large uncertainties. Apparent AGB-like colors may arise even if an object has already evolved into the post-AGB phase, depending on the spatial asymmetry of the circumstellar dust (e.g., torus-like structures) and the viewing angle. For instance, if a dense, cool equatorial envelope has formed due to a superwind, relatively warm components may still be visible along the polar direction. Additionally, some post-AGB stars are known to undergo temporary reactivation of mass loss (the so-called “born-again AGB” phenomenon).

Regarding the objects in category (2), we note their individual observational characteristics. IRAS 19229+1708 has been classified as an M-type supergiant of spectral type M3–-4~I based on its optical spectrum \citep{1994AAS...185.4515W}, and \citet{2021MNRAS.505.6051J} also identified it as a red supergiant (RSG), implying that it may be a relatively massive star. Typically, CSEs of spherically expanding AGB stars show double-peaked OH 1612~MHz maser line profiles \citep[][]{2004ApJS..155..595D}; however, both IRAS 19422+3506 and IRAS 22516+0838 exhibit multi-peaked OH 1612~MHz profiles \cite[for example,][]{2012A&A...537A...5W,2014AA...569A..92V}, suggesting possible deviations from a purely spherical CSE structure. Moreover, IRAS 19229+1708 shows a relatively large velocity excess of 34.3~km~s$^{-1}$ between the H$_2$O maser line and the satellite line, further supporting deviations from spherical expansion within the maser-emitting region.

Objects in category (1)—those with pulsation periods exceeding several hundred days and AGB-like colors—are almost certainly evolved stars at earlier evolutionary stages than category (3). If components deviating from spherical expansion exist near the centers of their CSEs, these phenomena would be highly intriguing in the context of the discussions in Sections~\ref{sec: intro} and \ref{sec: overall distribution in the IRAS colour-colour diagram}. So far, no WF exhibiting clear pulsation has been identified.

Finally, we note individual observational characteristics of the category (1) objects. IRAS 19068+1127 shows an OH 1612~MHz velocity width of 31.8~km~s$^{-1}$ (from $-$8.2~km~s$^{-1}$ to 23.6~km~s$^{-1}$) and an H$_2$O maser width of 35~km~s$^{-1}$ (from $-$13.0~km~s$^{-1}$ to 22.0~km~s$^{-1}$), with a blueshifted velocity excess of 4.8~km~s$^{-1}$ \citep{2024ApJS..270...13F}. Although a velocity excess is present, its magnitude is modest and insufficient to immediately classify the source as a WF candidate. In contrast, IRAS 19052+0922 shows an OH 1612~MHz width of 36.3~km~s$^{-1}$ (26.0–--62.3~km~s$^{-1}$) and an H$_2$O maser width of 49.2~km~s$^{-1}$ (5.6–--54.8~km~s$^{-1}$), with a blueshifted velocity excess of 20.4~km~s$^{-1}$. Considering that the typical expansion velocity of AGB CSEs is 10--–20~km~s$^{-1}$, this velocity excess is non-negligible and may indicate the presence of asymmetric motions in the central region of the CSE. Although this cannot be directly interpreted as a WF phenomenon, it may represent a precursor stage or “seed” of an extremely early WF phenomenon.

\section{Summary} \label{sec: summary}

In the present study, we systematically fitted the infrared SEDs of 17 CSEs, in which velocity excesses (non-spherical kinematic components) were identified by \citet{2024ApJS..270...13F} and \citet{2025ApJ...978..114X}, using the one-dimensional (spherically symmetric) dust radiative transfer model DUSTY. The aim was to search for objects whose outer CSE retains a spherical dust component, in order to identify WFs (or related phenomena) at the earliest evolutionary stages. The main conclusions are summarized as follows.

\begin{enumerate}
\item Among the 17 objects, good SED fits with DUSTY ($\chi^2 < 9.00\times10^{-4}$) were obtained for seven sources, suggesting that their outer CSEs are likely to be essentially spherical. These seven well-fitted sources comprise five single-peaked SED objects (IRAS 19052+0922, IRAS 19068+1127, IRAS 19229+1708, IRAS 19422+3506, IRAS 22516+0838) and two double-peaked SED objects (IRAS 17579$-$3121, IRAS 19319+2214). The remaining ten sources could not be adequately reproduced by the one-dimensional model, indicating that their CSEs likely exhibit significant non-spherical structures such as tori or jets.

\item In the IRAS two-color diagram, the objects that were well reproduced by the one-dimensional model predominantly exhibited AGB-like colors, whereas those that could not be reproduced primarily showed post-AGB-like colors. This trend is consistent with the morphological evolution from spherical CSEs (characteristic of the AGB phase) to non-spherical CSEs (characteristic of the post-AGB/PN phase). However, it was also suggested that when contributions from early-forming tori or small jets are minor, even CSEs containing such non-spherical components may be approximately reproduced by the one-dimensional model.

\item Based on infrared colors and variability properties, the seven well-fitted sources can be divided into three groups: (1) AGB-color objects exhibiting pulsation (IRAS 19052+0922, IRAS 19068+1127), (2) AGB-color objects without pulsation (IRAS 19229+1708, IRAS 19422+3506, IRAS 22516+0838), and (3) post-AGB-color objects without pulsation (IRAS 17579$-$3121, IRAS 19319+2214). Known WFs generally correspond to group (3). Groups (1) and (2) are likely at earlier evolutionary stages than known WFs and represent a promising parent population containing “very young WF candidates” that still retain spherical outer envelopes.

\item Overall, the results of the present study are consistent with a morphological evolution scenario in which a bipolar outflow is launched from the central region while the outer CSE remains spherical, and in which the resulting asymmetry subsequently propagates outward. In addition, considering the extent of the velocity excess together with other properties, the two objects IRAS 19229+1708 and IRAS 19052+0922 are inferred to be relatively plausible candidates for the earliest or incipient WF phase, in which asymmetric outflows begin to emerge while the outer envelope still retains a spherical morphology. High-angular-resolution imaging follow-up observations at radio and infrared wavelengths are anticipated.

\end{enumerate}

\begin{acknowledgments}
We acknowledge the science research grants from the China Manned Space Project with No. CMS-CSST-2021-A03, No.CMS-CSST-2021-B01. 
JN acknowledges financial support from the `One hundred top talent program of Sun Yat-sen University' grant no. 71000-18841229. YZ thanks the financial supports from the National Natural Science Foundation of China (NSFC, No.12473027 and No.12333005) and the Guangdong Basic and Applied Basic Research Funding (No.2024A1515010798)
\end{acknowledgments}

% \begin{acknowledgments}
% We acknowledge the science research grants from the China Manned Space Project with No. CMS-CSST-2021-A03, No.CMS-CSST-2021-B01. 
% JN acknowledges financial support from the `One hundred top talent program of Sun Yat-sen University' grant no. 71000-18841229. 
% \end{acknowledgments}

\appendix

\section{Two WFs not included in the SED analysis by Yung et al. (2017)}	\label{appendix: two WFs not included in the results of Yung et al. (2017)}

\begin{figure}
		\centering
  		\includegraphics[width=\textwidth, angle=0]{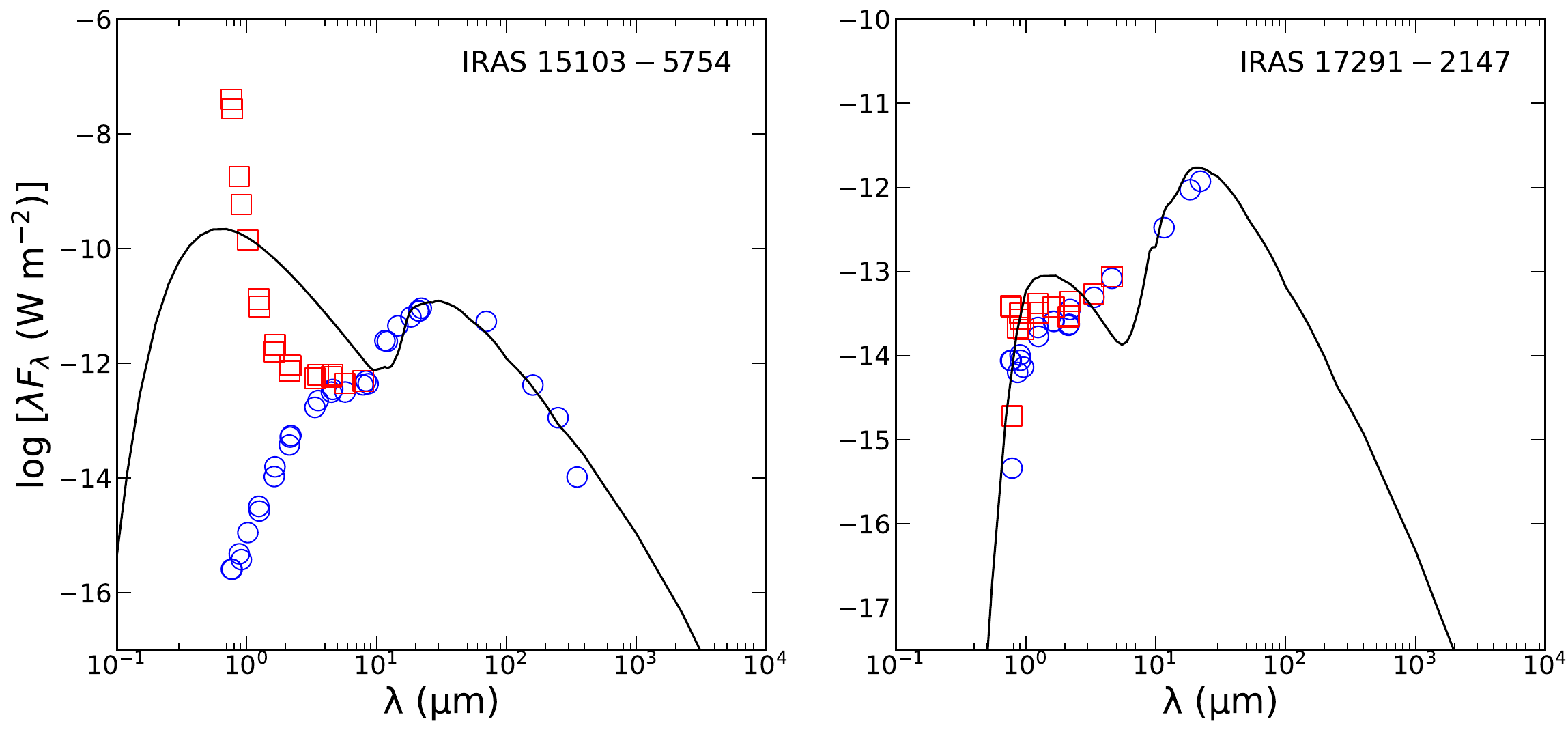}
\caption{Spectral energy distributions of two WFs (IRAS 15103--5754 and IRAS 17291--2147) that were not included in the SED analysis by \citet{2017MNRAS.465.4482Y}. The data points represent the observed photometry, and the curves show the results of the one-dimensional DUSTY model fittings performed in the present study. The plotting format is the same as that used in Figure~\ref{Fig: [spectral energy distributions of objects with velocity excess in the OH 1665/1667 MHz lines]}} \label{Fig: [spectral energy distributions of objects not included in Yung]}
\end{figure}

\citet{2017MNRAS.465.4482Y} performed one-dimensional DUSTY model fits to the SEDs of known WFs, following the same procedure as in the present study. However, of the 16 known WFs, two objects were not analyzed. These two WFs are IRAS 15103--5754 and IRAS 17291--2147. Therefore, in the present study, we performed one-dimensional DUSTY model fits for these two WFs to compare them with other WFs and WF candidates. The results of the fits are shown in Figure~\ref{Fig: [spectral energy distributions of objects not included in Yung]}. IRAS 15103--5754 is the only confirmed WF-type planetary nebula, whereas IRAS 17291--2147 was later identified as a bona fide WF by \citet{2017MNRAS.468.2081G} following the work of \citet{2017MNRAS.465.4482Y}. The procedures for obtaining the photometric data and for performing the one-dimensional DUSTY model fits were the same as those described in Section~\ref{sec: methodology and Data Origin}. The infrared SED data for both objects are also included, as for the other objects, in the electronic table attached to this paper. As shown in Figure~\ref{Fig: [spectral energy distributions of objects not included in Yung]}, no satisfactory fits were obtained for either case using the DUSTY model.

% \section{Two WFs not included in the results of Yung et al. (2017)}	\label{appendix: two WFs not included in the results of Yung et al. (2017)}

% Figure~\ref{Fig: [spectral energy distributions of objects not included in Yung]} shows the spectral energy distributions of two WFs (IRAS 15103--5754 and IRAS 17291--2147) who are not included in the results of \citet{2017MNRAS.465.4482Y}. IRAS 15103--5754 was the only confirmed WF PN, and IRAS 17291--2147 was recognized as a bona fide WF in \citet{2017MNRAS.468.2081G}, following the work of \citet{2017MNRAS.465.4482Y}. The procedures for acquiring and analyzing the photometric data were identical to those for the target objects and are described in Section~\ref{sec: methodology and Data Origin}. Both two cases do not have a good fit from DUSTY models. Their infrared SED data are available in the electronic table attached to this paper as well.

%%%%%%%%%%%%figures
\clearpage
%-------------------------------------------------------------
  % \FloatBarrier
 %\afterpage{
\clearpage

  % \FloatBarrier
 %\afterpage{
\clearpage

%%%%%%%%%%%%%%%%%%%%positions of good-fitting cases and no good-fitting cases
\clearpage

\clearpage

\clearpage
\FloatBarrier
 
\clearpage

%%%%apppendix
\clearpage

\clearpage
%%%%%%%%%%%%tables
\input{SED_data}
\input{Parameters_used_in_generating_the_DUSTY_codes}

\input{Objects_used_in_this_study}
\input{IRAS_Variablitiy_WISE_light_curve}

\bibliography{ms}{}
\bibliographystyle{aasjournal}

\end{CJK*}
\end{document}

%% file: SED_data.tex
% % \clearpage
% % \startlongtable
% \begin{deluxetable}{ccc} 
% \tablecaption{Descriptive Table of the Infrared Photometric Data.}
% \label{Tab: [SED data]}
% \tablehead{
% 			Column & Unit/Format & Description }
% \startdata
% Source Name & & Name of the object\\
% Band Name &   &Name of the photometric band\\
% Wavelength & $\rm \mu m$ & Wavelength of the photometric band\\
% Flux &$\rm W~m^{-2}$ & Flux of the photometric band \\
% Flux Error &$\rm W~m^{-2}$ &Mean error/standard deviations of flux\\
% R.A.(J2000) &hh mm ss.ss & Right ascension of the object\\
% Dec.(J2000) &dd mm ss.ss & Declination of the object\\
% \enddata
% \tablecomments{In the electronic catalog, these values are provided in separate blocks for each object.}
% \end{deluxetable}

% \clearpage
% \startlongtable
\begin{deluxetable}{ccc} 
\tablecaption{Descriptive Table of the Infrared Photometric Data.}
\label{Tab: [SED data]}
\tablehead{
			Column & Unit/Format & Description }
\startdata
Source Name & & Name of the object\\
Band Name &   &Name of the photometric band\\
Wavelength & $\rm \mu m$ & Wavelength of the photometric band\\
Flux &$\rm W~m^{-2}$ & Integrated flux in the photometric band \\
Flux Error &$\rm W~m^{-2}$ &Standard deviation of the flux\\
R.A.(J2000) &hh mm ss.ss & Right ascension of the object\\
Dec.(J2000) &dd mm ss.ss & Declination of the object\\
\enddata
\tablecomments{In the electronic catalog, these values are provided in separate entries for each object.}
\end{deluxetable}

%% file: Parameters_used_in_generating_the_DUSTY_codes.tex
% \begin{deluxetable}{cccc} 
% \caption{The Models Generated by DUSTY Codes in Different Parameters.}
% \label{Tab: [The_models_generated_by_DUSTY_codes_in_different_parameters]}
% \tablehead{	\multirow{1}{*}{Physical Parameter}  & \multicolumn{1}{c}{Range} & \multicolumn{1}{c}{Step} &\multicolumn{1}{c}{Number}}
% \startdata
% $T_{\rm eff}$ (K) & 1400--10000  & 200 & 44\\
% $T_{\rm d}$ (K) & 50--1500 & 50 & 30\\
% $\tau_{\rm 2.2}$ (low range)  & 0.01--1 & 0.0168 &60 \\
% $\tau_{\rm 2.2}$  (high range) & 1.5--20 & 0.5 & 38\\
% \enddata
% \end{deluxetable}

\begin{deluxetable}{cccc} 
\caption{Parameter Ranges for the DUSTY Model Grid.}
\label{Tab: [The_models_generated_by_DUSTY_codes_in_different_parameters]}
\tablehead{	\multirow{1}{*}{Physical Parameter}  & \multicolumn{1}{c}{Range} & \multicolumn{1}{c}{Step Size} &\multicolumn{1}{c}{Number of Steps}}
\startdata
$T_{\rm eff}$ (K) & 1400 to 10000  & 200 & 44\\
$T_{\rm d}$ (K) & 50 to 1500 & 50 & 30\\
$\tau_{\rm 2.2}$ (low range)  & 0.01 to 1 & 0.0168 &60 \\
$\tau_{\rm 2.2}$  (high range) & 1.5 to 20 & 0.5 & 38\\
\enddata
\tablecomments{Number of Steps refers to the number of discrete grid points for each parameter.}
\end{deluxetable}

%% file: Objects_used_in_this_study.tex
\begin{deluxetable}{ccccccccccc} 
\caption{Basic Information and Best-Fit DUSTY Model Parameters of the Target Objects.}\label{Tab: [list of objects used in this study and Dusty parameters]}
\tablehead{	 \multirow{1}{*}{Source name}  & \multicolumn{1}{c}{R.A.} & \multicolumn{1}{c}{Dec.} &\multicolumn{1}{c}{$A_{\rm V}$} & \multicolumn{1}{c}{DUSTY Profile} & \multicolumn{1}{c}{$T_{\rm eff}$} &\multicolumn{1}{c}{$T_{\rm d}$} & \multicolumn{1}{c}{$\tau_{\rm 2.2}$} &\multicolumn{1}{c}{$\dot{M}$} &\multicolumn{1}{c}{$M_{\rm star}$}&  \multicolumn{1}{c}{$\chi^2$}\\ 
&(J2000)&(J2000)  &&&(K) &(K) & &(M$_{\odot}$~yr$^{-1}$) &M$_{\odot}$ }
\startdata
\multicolumn{11}{c}{\textit{From \citet{2024ApJS..270...13F}}}\\
\hline
IRAS 18588$+$0428 &19 01 20.0  &+04 32 31.3  & 66.526 &... &... &... &...&...&...&1.13$\times 10^{-1}$\\
IRAS 19052$+$0922 &19 07 39.1 &+09 27 15.8  &11.849 &S &4400 & 850 & 0.883&$2.02\times10^{-5}$&$<$4.34&4.19$\times 10^{-4}$\\
IRAS 19067$+$0811 &19 09 08.3 &+08 16 34.0   &48.891 &...&... &... &...&...&...&2.00$\times 10^{-2}$\\
IRAS 19068$+$1127 &19 09 11.7 &+11 32 43.0  &6.431 &S &5000 &1250 &0.849&$1.38\times10^{-5}$& $<$4.78&3.91$\times 10^{-4}$\\
IRAS 19083$+$0851 &19 10 47.3 &+08 56 22.6  &29.088 &...  &... &... &...&...&...&1.25$\times 10^{-2}$\\
IRAS 19229$+$1708 &19 25 12.5 &+17 14 49.2  &14.145 &S &6200 & 1250 &0.010&$4.02\times10^{-7}$& $<$3.78&6.54$\times 10^{-4}$\\
IRAS 19319$+$2214 &19 34 03.6 &+22 21 16.0  & 9.493 &D &6600 & 500 & 0.648&$3.35\times10^{-5}$& $<$6.06&4.67$\times 10^{-4}$\\
IRAS 19422$+$3506 &19 44 07.0 &+35 14 08.1  &0.802 &S &10000 & 900 & 1.000&$2.79\times10^{-5}$& $<$4.30&6.11$\times 10^{-4}$\\
IRAS 22516$+$0838 &22 54 11.4 &+08 54 07.2  &0.180 &S &2000 & 900 & 0.429&$5.61\times10^{-6}$ & $<$3.42&4.58$\times 10^{-4}$\\
\hline
\multicolumn{11}{c}{\textit{From \citet{2025ApJ...978..114X}}}\\
\hline
IRAS 15405$-$4945 &15 44 11.1 &$-$49 55 20.0  &3.248 &... &... &... &...&...&...&2.66$\times 10^{-3}$\\%%
IRAS 17579$-$3121 &18 01 13.4 &$-$31 21 56.0  &2.189 &D  &6600&100&0.228 &$1.08\times10^{-4}$ & $<$1.30&3.24$\times 10^{-4}$\\
IRAS 18251$-$1048 & 18 27 56.4 &$-$10 46 54.1  &32.972 &... &... &... &...&...&...&3.53$\times 10^{-3}$\\
IRAS 18498$-$0017 &18 52 22.2 &$-$00 14 11.0 &35.739 &... &... &... &...&...&...&2.36$\times 10^{-2}$\\%%
IRAS 19069$+$0916 &19 09 19.2 &+09 21 11.3  &14.160 &...&... &... &...&...&...&2.41$\times 10^{-3}$\\
IRAS 19103$+$0913 &19 12 42.4 &+09 24 26.0  &18.296 &...&... &... &...&...&...&1.12$\times 10^{-3}$\\%
IRAS 19190$+$1102 &19 21 25.2 &+11 08 40.0  &9.464 &...&...  &... &...&...&...&3.10$\times 10^{-3}$\\%
IRAS 19352$+$2030 &19 37 24.0 &+20 36 58.0  &18.900 &...&... &... &...&...&...&5.77$\times 10^{-3}$\\%%
\enddata
\tablecomments{The “DUSTY Profile” column indicates the shape of the fitted model profile: ``S'' denotes a single-peaked profile, while ``D'' indicates a double-peaked one.  
$\dot{M}$ represents the mass-loss rate.  
For sources without a reliable model fit (i.e., no solution), the corresponding entries are indicated by “...”.}
\end{deluxetable}

%% file: IRAS_Variablitiy_WISE_light_curve.tex
\begin{deluxetable*}{cccc}
\caption{IRAS Variability Index and WISE Light Curves of the Target Sources.}
\label{Tab: [IRAS variability index and WISE light curve for our target objects]}
\tablehead{ \multicolumn{1}{c}{Source name}	& \multicolumn{1}{c}{IRAS Var. Index (\%)}  & \multicolumn{1}{c}{WISE Periodicity} &\multicolumn{1}{c}{WISE Period (days)} }
\startdata
\multicolumn{4}{c}{Well-fitted sources} \\ 
\hline
IRAS 17579$-$3121$^{\dagger}$ & 3 &Non-periodic &...\\
IRAS 19052$+$0922 & 99 & Periodic &644.7  \\
IRAS 19068$+$1127 & 99 & Periodic & 662.9  \\
IRAS 19229$+$1708 &7 &...&no data\\
IRAS 19319$+$2214 & 3 & Non-periodic &...\\
IRAS 19422$+$3506 & 8 &...&no data\\
IRAS 22516$+$0838 & 0 &...&no data\\
\hline
\multicolumn{4}{c}{Poorly-fitted sources} \\ 
\hline
IRAS 15405$-$4945$^{\dagger}$ &$-1$ &Non-periodic&...\\%%
IRAS 18251$-$1048$^{\dagger}$ & 72 & Periodic &1396.5\\
IRAS 18498$-$0017$^{\dagger}$ & 96 & Periodic &1583.2\\%%
IRAS 18588$+$0428 & 94 & Non-periodic&...\\
IRAS 19067$+$0811 & 98 & Periodic & 2008.0 \\
IRAS 19069$+$0916$^{\dagger}$ & 36 & Periodic &1384.4 \\
IRAS 19083$+$0851 & 77 &Non-periodic&...\\
IRAS 19103$+$0913$^{\dagger}$ & 99 & Periodic &505.6 \\%
IRAS 19190$+$1102$^{\dagger}$ & 9  &Non-periodic&...\\%
IRAS 19352$+$2030$^{\dagger}$ & 99 & Periodic &517.5  \\%%
\enddata
\tablecomments{IRAS Var. Index: Percent probability (0--99\%) that a source is variable, derived from the 12 and 25 micron flux densities and their uncertainties; “$-1$” denotes not examined.
WISE Periodicity: “...” denotes too few data points to construct a WISE light curve. WISE Period (days): “...” denotes no detected periodicity. WISE light curves of the \citet{2024ApJS..270...13F} objects are shown in Figure~\ref{Fig: [WISE light curves with periodicity for the sources of Fan et al.2024]} and Figure~\ref{Fig: [WISE light curves of known WFs]}; additional curves for sources marked $^{\dagger}$ appear in \citet{2025ApJ...978..114X}.}
\end{deluxetable*}

%% file: ms.bib
@INPROCEEDINGS{2004evn..conf..169A,
       author = {{Alcolea}, J.},
        title = "{Circumstellar masers}",
    booktitle = {European VLBI Network on New Developments in VLBI Science and Technology},
         year = 2004,
       editor = {{Bachiller}, R. and {Colomer}, F. and {Desmurs}, J. -F. and {de Vicente}, P.},
        month = jan,
        pages = {169-176},
          doi = {10.48550/arXiv.astro-ph/0412295},
archivePrefix = {arXiv},
       eprint = {astro-ph/0412295},
 primaryClass = {astro-ph},
       adsurl = {https://ui.adsabs.harvard.edu/abs/2004evn..conf..169A}
}

@article{Alam_2015,
doi = {10.1088/0067-0049/219/1/12},
url = {https://dx.doi.org/10.1088/0067-0049/219/1/12},
year = {2015},
month = {jul},
publisher = {The American Astronomical Society},
volume = {219},
number = {1},
pages = {12},
author = {Shadab Alam and Franco D. Albareti and Carlos Allende Prieto and F. Anders and Scott F. Anderson and Timothy Anderton and Brett H. Andrews and Eric Armengaud and Éric Aubourg and Stephen Bailey and Sarbani Basu and Julian E. Bautista and Rachael L. Beaton and Timothy C. Beers and Chad F. Bender and Andreas A. Berlind and Florian Beutler and Vaishali Bhardwaj and Jonathan C. Bird and Dmitry Bizyaev and Cullen H. Blake and Michael R. Blanton and Michael Blomqvist and John J. Bochanski and Adam S. Bolton and Jo Bovy and A. Shelden Bradley and W. N. Brandt and D. E. Brauer and J. Brinkmann and Peter J. Brown and Joel R. Brownstein and Angela Burden and Etienne Burtin and Nicolás G. Busca and Zheng Cai and Diego Capozzi and Aurelio Carnero Rosell and Michael A. Carr and Ricardo Carrera and K. C. Chambers and William James Chaplin and Yen-Chi Chen and Cristina Chiappini and S. Drew Chojnowski and Chia-Hsun Chuang and Nicolas Clerc and Johan Comparat and Kevin Covey and Rupert A. C. Croft and Antonio J. Cuesta and Katia Cunha and Luiz N. da Costa and Nicola Da Rio and James R. A. Davenport and Kyle S. Dawson and Nathan De Lee and Timothée Delubac and Rohit Deshpande and Saurav Dhital and Letícia Dutra-Ferreira and Tom Dwelly and Anne Ealet and Garrett L. Ebelke and Edward M. Edmondson and Daniel J. Eisenstein and Tristan Ellsworth and Yvonne Elsworth and Courtney R. Epstein and Michael Eracleous and Stephanie Escoffier and Massimiliano Esposito and Michael L. Evans and Xiaohui Fan and Emma Fernández-Alvar and Diane Feuillet and Nurten Filiz Ak and Hayley Finley and Alexis Finoguenov and Kevin Flaherty and Scott W. Fleming and Andreu Font-Ribera and Jonathan Foster and Peter M. Frinchaboy and J. G. Galbraith-Frew and Rafael A. García and D. A. García-Hernández and Ana E. García Pérez and Patrick Gaulme and Jian Ge and R. Génova-Santos and A. Georgakakis and Luan Ghezzi and Bruce A. Gillespie and Léo Girardi and Daniel Goddard and Satya Gontcho A Gontcho and Jonay I. González Hernández and Eva K. Grebel and Paul J. Green and Jan Niklas Grieb and Nolan Grieves and James E. Gunn and Hong Guo and Paul Harding and Sten Hasselquist and Suzanne L. Hawley and Michael Hayden and Fred R. Hearty and Saskia Hekker and Shirley Ho and David W. Hogg and Kelly Holley-Bockelmann and Jon A. Holtzman and Klaus Honscheid and Daniel Huber and Joseph Huehnerhoff and Inese I. Ivans and Linhua Jiang and Jennifer A. Johnson and Karen Kinemuchi and David Kirkby and Francisco Kitaura and Mark A. Klaene and Gillian R. Knapp and Jean-Paul Kneib and Xavier P. Koenig and Charles R. Lam and Ting-Wen Lan and Dustin Lang and Pierre Laurent and Jean-Marc Le Goff and Alexie Leauthaud and Khee-Gan Lee and Young Sun Lee and Timothy C. Licquia and Jian Liu and Daniel C. Long and Martín López-Corredoira and Diego Lorenzo-Oliveira and Sara Lucatello and Britt Lundgren and Robert H. Lupton and Claude E. Mack III and Suvrath Mahadevan and Marcio A. G. Maia and Steven R. Majewski and Elena Malanushenko and Viktor Malanushenko and A. Manchado and Marc Manera and Qingqing Mao and Claudia Maraston and Robert C. Marchwinski and Daniel Margala and Sarah L. Martell and Marie Martig and Karen L. Masters and Savita Mathur and Cameron K. McBride and Peregrine M. McGehee and Ian D. McGreer and Richard G. McMahon and Brice Ménard and Marie-Luise Menzel and Andrea Merloni and Szabolcs Mészáros and Adam A. Miller and Jordi Miralda-Escudé and Hironao Miyatake and Antonio D. Montero-Dorta and Surhud More and Eric Morganson and Xan Morice-Atkinson and Heather L. Morrison and Benôit Mosser and Demitri Muna and Adam D. Myers and Kirpal Nandra and Jeffrey A. Newman and Mark Neyrinck and Duy Cuong Nguyen and Robert C. Nichol and David L. Nidever and Pasquier Noterdaeme and Sebastián E. Nuza and Julia E. O’Connell and Robert W. O’Connell and Ross O’Connell and Ricardo L. C. Ogando and Matthew D. Olmstead and Audrey E. Oravetz and Daniel J. Oravetz and Keisuke Osumi and Russell Owen and Deborah L. Padgett and Nikhil Padmanabhan and Martin Paegert and Nathalie Palanque-Delabrouille and Kaike Pan and John K. Parejko and Isabelle Pâris and Changbom Park and Petchara Pattarakijwanich and M. Pellejero-Ibanez and Joshua Pepper and Will J. Percival and Ismael Pérez-Fournon and Ignasi Pe´rez-Ra`fols and Patrick Petitjean and Matthew M. Pieri and Marc H. Pinsonneault and Gustavo F. Porto de Mello and Francisco Prada and Abhishek Prakash and Adrian M. Price-Whelan and Pavlos Protopapas and M. Jordan Raddick and Mubdi Rahman and Beth A. Reid and James Rich and Hans-Walter Rix and Annie C. Robin and Constance M. Rockosi and Thaíse S. Rodrigues and Sergio Rodríguez-Torres and Natalie A. Roe and Ashley J. Ross and Nicholas P. Ross and Graziano Rossi and John J. Ruan and J. A. Rubiño-Martín and Eli S. Rykoff and Salvador Salazar-Albornoz and Mara Salvato and Lado Samushia and Ariel G. Sánchez and Basílio Santiago and Conor Sayres and Ricardo P. Schiavon and David J. Schlegel and Sarah J. Schmidt and Donald P. Schneider and Mathias Schultheis and Axel D. Schwope and C. G. Scóccola and Caroline Scott and Kris Sellgren and Hee-Jong Seo and Aldo Serenelli and Neville Shane and Yue Shen and Matthew Shetrone and Yiping Shu and V. Silva Aguirre and Thirupathi Sivarani and M. F. Skrutskie and Anže Slosar and Verne V. Smith and Flávia Sobreira and Diogo Souto and Keivan G. Stassun and Matthias Steinmetz and Dennis Stello and Michael A. Strauss and Alina Streblyanska and Nao Suzuki and Molly E. C. Swanson and Jonathan C. Tan and Jamie Tayar and Ryan C. Terrien and Aniruddha R. Thakar and Daniel Thomas and Neil Thomas and Benjamin A. Thompson and Jeremy L. Tinker and Rita Tojeiro and Nicholas W. Troup and Mariana Vargas-Magaña and Jose A. Vazquez and Licia Verde and Matteo Viel and Nicole P. Vogt and David A. Wake and Ji Wang and Benjamin A. Weaver and David H. Weinberg and Benjamin J. Weiner and Martin White and John C. Wilson and John P. Wisniewski and W. M. Wood-Vasey and Christophe Ye`che and Donald G. York and Nadia L. Zakamska and O. Zamora and Gail Zasowski and Idit Zehavi and Gong-Bo Zhao and Zheng Zheng and Xu Zhou (周旭) and Zhimin Zhou (周志民) and Hu Zou (邹虎) and Guangtun Zhu},
title = {THE ELEVENTH AND TWELFTH DATA RELEASES OF THE SLOAN DIGITAL SKY SURVEY: FINAL DATA FROM SDSS-III},
journal = {The Astrophysical Journal Supplement Series}
}

@PROCEEDINGS{1988iras....1.....B,
        title = "{Infrared Astronomical Satellite (IRAS) Catalogs and Atlases.Volume 1: Explanatory Supplement.}",
    booktitle = {Infrared astronomical satellite (IRAS) catalogs and atlases. Volume 1: Explanatory supplement},
         year = 1988,
       editor = {{Beichman}, C.~A. and {Neugebauer}, G. and {Habing}, H.~J. and {Clegg}, P.~E. and {Chester}, Thomas J.},
       volume = {1},
        month = jan,
       adsurl = {https://ui.adsabs.harvard.edu/abs/1988iras....1.....B}
}

@ARTICLE{2003PASP..115..953B,
       author = {{Benjamin}, Robert A. and {Churchwell}, E. and {Babler}, Brian L. and {Bania}, T.~M. and {Clemens}, Dan P. and {Cohen}, Martin and {Dickey}, John M. and {Indebetouw}, R{\'e}my and {Jackson}, James M. and {Kobulnicky}, Henry A. and {Lazarian}, Alex and {Marston}, A.~P. and {Mathis}, John S. and {Meade}, Marilyn R. and {Seager}, Sara and {Stolovy}, S.~R. and {Watson}, C. and {Whitney}, Barbara A. and {Wolff}, Michael J. and {Wolfire}, Mark G.},
        title = "{GLIMPSE. I. An SIRTF Legacy Project to Map the Inner Galaxy}",
      journal = {\pasp},
         year = 2003,
        month = aug,
       volume = {115},
       number = {810},
        pages = {953-964},
          doi = {10.1086/376696},
archivePrefix = {arXiv},
       eprint = {astro-ph/0306274},
 primaryClass = {astro-ph},
       adsurl = {https://ui.adsabs.harvard.edu/abs/2003PASP..115..953B}
}

@INPROCEEDINGS{2018AAS...23110201C,
       author = {{Chambers}, Kenneth and {Pan-STARRS Team}},
        title = "{The Pan-STARRS1 Surveys}",
    booktitle = {American Astronomical Society Meeting Abstracts \#231},
         year = 2018,
       series = {American Astronomical Society Meeting Abstracts},
       volume = {231},
        month = jan,
          eid = {102.01},
        pages = {102.01},
       adsurl = {https://ui.adsabs.harvard.edu/abs/2018AAS...23110201C}
}

@ARTICLE{1989RPPh...52..881C,
       author = {{Cohen}, R.~J.},
        title = "{Compact maser sources}",
      journal = {Reports on Progress in Physics},
         year = 1989,
        month = aug,
       volume = {52},
       number = {8},
        pages = {881-943},
          doi = {10.1088/0034-4885/52/8/001},
       adsurl = {https://ui.adsabs.harvard.edu/abs/1989RPPh...52..881C}
}

@ARTICLE{2012RAA....12.1197C,
       author = {{Cui}, Xiang-Qun and {Zhao}, Yong-Heng and {Chu}, Yao-Quan and {Li}, Guo-Ping and {Li}, Qi and {Zhang}, Li-Ping and {Su}, Hong-Jun and {Yao}, Zheng-Qiu and {Wang}, Ya-Nan and {Xing}, Xiao-Zheng and {Li}, Xin-Nan and {Zhu}, Yong-Tian and {Wang}, Gang and {Gu}, Bo-Zhong and {Luo}, A. -Li and {Xu}, Xin-Qi and {Zhang}, Zhen-Chao and {Liu}, Gen-Rong and {Zhang}, Hao-Tong and {Yang}, De-Hua and {Cao}, Shu-Yun and {Chen}, Hai-Yuan and {Chen}, Jian-Jun and {Chen}, Kun-Xin and {Chen}, Ying and {Chu}, Jia-Ru and {Feng}, Lei and {Gong}, Xue-Fei and {Hou}, Yong-Hui and {Hu}, Hong-Zhuan and {Hu}, Ning-Sheng and {Hu}, Zhong-Wen and {Jia}, Lei and {Jiang}, Fang-Hua and {Jiang}, Xiang and {Jiang}, Zi-Bo and {Jin}, Ge and {Li}, Ai-Hua and {Li}, Yan and {Li}, Ye-Ping and {Liu}, Guan-Qun and {Liu}, Zhi-Gang and {Lu}, Wen-Zhi and {Mao}, Yin-Dun and {Men}, Li and {Qi}, Yong-Jun and {Qi}, Zhao-Xiang and {Shi}, Huo-Ming and {Tang}, Zheng-Hong and {Tao}, Qing-Sheng and {Wang}, Da-Qi and {Wang}, Dan and {Wang}, Guo-Min and {Wang}, Hai and {Wang}, Jia-Ning and {Wang}, Jian and {Wang}, Jian-Ling and {Wang}, Jian-Ping and {Wang}, Lei and {Wang}, Shu-Qing and {Wang}, You and {Wang}, Yue-Fei and {Xu}, Ling-Zhe and {Xu}, Yan and {Yang}, Shi-Hai and {Yu}, Yong and {Yuan}, Hui and {Yuan}, Xiang-Yan and {Zhai}, Chao and {Zhang}, Jing and {Zhang}, Yan-Xia and {Zhang}, Yong and {Zhao}, Ming and {Zhou}, Fang and {Zhou}, Guo-Hua and {Zhu}, Jie and {Zou}, Si-Cheng},
        title = "{The Large Sky Area Multi-Object Fiber Spectroscopic Telescope (LAMOST)}",
      journal = {Research in Astronomy and Astrophysics},
         year = 2012,
        month = sep,
       volume = {12},
       number = {9},
        pages = {1197-1242},
          doi = {10.1088/1674-4527/12/9/003},
       adsurl = {https://ui.adsabs.harvard.edu/abs/2012RAA....12.1197C}
}

@INPROCEEDINGS{2012IAUS..287..217D,
       author = {{Desmurs}, J. -F.},
        title = "{Maser emission during post-AGB evolution}",
    booktitle = {Cosmic Masers - from OH to H0},
         year = 2012,
       editor = {{Booth}, Roy S. and {Vlemmings}, Wouter H.~T. and {Humphreys}, Elizabeth M.~L.},
       series = {IAU Symposium},
       volume = {287},
        month = jul,
        pages = {217-224},
          doi = {10.1017/S1743921312006990},
archivePrefix = {arXiv},
       eprint = {1210.4373},
 primaryClass = {astro-ph.GA},
       adsurl = {https://ui.adsabs.harvard.edu/abs/2012IAUS..287..217D}
}

@ARTICLE{1994ApJ...430L..61D,
       author = {{Diamond}, P.~J. and {Kemball}, A.~J. and {Junor}, W. and {Zensus}, A. and {Benson}, J. and {Dhawan}, V.},
        title = "{Observation of a Ring Structure in SiO Maser Emission from Late-Type Stars}",
      journal = {\apjl},
         year = 1994,
        month = jul,
       volume = {430},
        pages = {L61},
          doi = {10.1086/187438},
       adsurl = {https://ui.adsabs.harvard.edu/abs/1994ApJ...430L..61D}
}

@ARTICLE{1984ApJ...285...89D,
       author = {{Draine}, B.~T. and {Lee}, H.~M.},
        title = "{Optical Properties of Interstellar Graphite and Silicate Grains}",
      journal = {\apj},
         year = 1984,
        month = oct,
       volume = {285},
        pages = {89},
          doi = {10.1086/162480},
       adsurl = {https://ui.adsabs.harvard.edu/abs/1984ApJ...285...89D}
}

@ARTICLE{2004ApJS..155..595D,
       author = {{Deacon}, R.~M. and {Chapman}, J.~M. and {Green}, A.~J.},
        title = "{OH Maser Observations of Likely Planetary Nebulae Precursors}",
      journal = {\apjs},
         year = 2004,
        month = dec,
       volume = {155},
       number = {2},
        pages = {595-622},
          doi = {10.1086/425329},
archivePrefix = {arXiv},
       eprint = {astro-ph/0406198},
 primaryClass = {astro-ph},
       adsurl = {https://ui.adsabs.harvard.edu/abs/2004ApJS..155..595D}
}

@ARTICLE{1996A&A...313..605D,
       author = {{Dorfi}, E.~A. and {Hoefner}, S.},
        title = "{Non-spherical dust driven winds of slowly rotating AGB stars.}",
      journal = {\aap},
         year = 1996,
        month = sep,
       volume = {313},
        pages = {605-610},
       adsurl = {https://ui.adsabs.harvard.edu/abs/1996A&A...313..605D}
}

@ARTICLE{2025MNRAS.539.1220D,
       author = {{Dai}, Min and {Wang}, Shu and {Jiang}, Biwei},
        title = "{The binary fraction of red supergiants in the Magellanic Clouds}",
      journal = {\mnras},
         year = 2025,
        month = may,
       volume = {539},
       number = {2},
        pages = {1220-1235},
          doi = {10.1093/mnras/staf560},
archivePrefix = {arXiv},
       eprint = {2504.03357},
 primaryClass = {astro-ph.SR},
       adsurl = {https://ui.adsabs.harvard.edu/abs/2025MNRAS.539.1220D}
}

@ARTICLE{2017MNRAS.471..100E,
       author = {{Elia}, Davide and {Molinari}, S. and {Schisano}, E. and {Pestalozzi}, M. and {Pezzuto}, S. and {Merello}, M. and {Noriega-Crespo}, A. and {Moore}, T.~J.~T. and {Russeil}, D. and {Mottram}, J.~C. and {Paladini}, R. and {Strafella}, F. and {Benedettini}, M. and {Bernard}, J.~P. and {Di Giorgio}, A. and {Eden}, D.~J. and {Fukui}, Y. and {Plume}, R. and {Bally}, J. and {Martin}, P.~G. and {Ragan}, S.~E. and {Jaffa}, S.~E. and {Motte}, F. and {Olmi}, L. and {Schneider}, N. and {Testi}, L. and {Wyrowski}, F. and {Zavagno}, A. and {Calzoletti}, L. and {Faustini}, F. and {Natoli}, P. and {Palmeirim}, P. and {Piacentini}, F. and {Piazzo}, L. and {Pilbratt}, G.~L. and {Polychroni}, D. and {Baldeschi}, A. and {Beltr{\'a}n}, M.~T. and {Billot}, N. and {Cambr{\'e}sy}, L. and {Cesaroni}, R. and {Garc{\'\i}a-Lario}, P. and {Hoare}, M.~G. and {Huang}, M. and {Joncas}, G. and {Liu}, S.~J. and {Maiolo}, B.~M.~T. and {Marsh}, K.~A. and {Maruccia}, Y. and {M{\`e}ge}, P. and {Peretto}, N. and {Rygl}, K.~L.~J. and {Schilke}, P. and {Thompson}, M.~A. and {Traficante}, A. and {Umana}, G. and {Veneziani}, M. and {Ward-Thompson}, D. and {Whitworth}, A.~P. and {Arab}, H. and {Bandieramonte}, M. and {Becciani}, U. and {Brescia}, M. and {Buemi}, C. and {Bufano}, F. and {Butora}, R. and {Cavuoti}, S. and {Costa}, A. and {Fiorellino}, E. and {Hajnal}, A. and {Hayakawa}, T. and {Kacsuk}, P. and {Leto}, P. and {Li Causi}, G. and {Marchili}, N. and {Martinavarro-Armengol}, S. and {Mercurio}, A. and {Molinaro}, M. and {Riccio}, G. and {Sano}, H. and {Sciacca}, E. and {Tachihara}, K. and {Torii}, K. and {Trigilio}, C. and {Vitello}, F. and {Yamamoto}, H.},
        title = "{The Hi-GAL compact source catalogue - I. The physical properties of the clumps in the inner Galaxy (-71.0{\textdegree} < {\ensuremath{\ell}} < 67.0{\textdegree})}",
      journal = {\mnras},
         year = 2017,
        month = oct,
       volume = {471},
       number = {1},
        pages = {100-143},
          doi = {10.1093/mnras/stx1357},
archivePrefix = {arXiv},
       eprint = {1706.01046},
 primaryClass = {astro-ph.GA},
       adsurl = {https://ui.adsabs.harvard.edu/abs/2017MNRAS.471..100E}
}

@ARTICLE{2006Msngr.126...41E,
       author = {{Emerson}, J. and {McPherson}, A. and {Sutherland}, W.},
        title = "{Visible and Infrared Survey Telescope for Astronomy: Progress Report}",
      journal = {The Messenger},
         year = 2006,
        month = dec,
       volume = {126},
        pages = {41-42},
       adsurl = {https://ui.adsabs.harvard.edu/abs/2006Msngr.126...41E}
}

@INPROCEEDINGS{2003AAS...203.5708E,
       author = {{Egan}, M.~P. and {Price}, S.~D. and {Kraemer}, K.~E.},
        title = "{The Midcourse Space Experiment Point Source Catalog Version 2.3}",
    booktitle = {American Astronomical Society Meeting Abstracts},
         year = 2003,
       series = {American Astronomical Society Meeting Abstracts},
       volume = {203},
        month = dec,
          eid = {57.08},
        pages = {57.08},
       adsurl = {https://ui.adsabs.harvard.edu/abs/2003AAS...203.5708E}
}

@ARTICLE{2002A&A...388..252E,
       author = {{Engels}, D.},
        title = "{Water vapor masers in stars departing from the AGB}",
      journal = {\aap},
         year = 2002,
        month = jun,
       volume = {388},
        pages = {252-267},
          doi = {10.1051/0004-6361:20020483},
       adsurl = {https://ui.adsabs.harvard.edu/abs/2002A&A...388..252E}
}

@ARTICLE{2024ApJS..270...13F,
       author = {{Fan}, Haichen and {Nakashima}, Jun-ichi and {Engels}, D. and {Zhang}, Yong and {Qiu}, Jian-Jie and {Feng}, Huan-Xue and {Xie}, Jia-Yong and {Imai}, Hiroshi and {Hsia}, Chih-Hao},
        title = "{Systematic Search for Water Fountain Candidates Using the Databases of Circumstellar Maser Sources}",
      journal = {\apjs},
         year = 2024,
        month = jan,
       volume = {270},
       number = {1},
          eid = {13},
        pages = {13},
          doi = {10.3847/1538-4365/ad0458},
archivePrefix = {arXiv},
       eprint = {2310.05065},
 primaryClass = {astro-ph.SR},
       adsurl = {https://ui.adsabs.harvard.edu/abs/2024ApJS..270...13F}
}

@ARTICLE{2016A&A...595A...1G,
       author = {{Gaia Collaboration} and {Prusti}, T. and {de Bruijne}, J.~H.~J. and {Brown}, A.~G.~A. and {Vallenari}, A. and {Babusiaux}, C. and {Bailer-Jones}, C.~A.~L. and {Bastian}, U. and {Biermann}, M. and {Evans}, D.~W. and {Eyer}, L. and {Jansen}, F. and {Jordi}, C. and {Klioner}, S.~A. and {Lammers}, U. and {Lindegren}, L. and {Luri}, X. and {Mignard}, F. and {Milligan}, D.~J. and {Panem}, C. and {Poinsignon}, V. and {Pourbaix}, D. and {Randich}, S. and {Sarri}, G. and {Sartoretti}, P. and {Siddiqui}, H.~I. and {Soubiran}, C. and {Valette}, V. and {van Leeuwen}, F. and {Walton}, N.~A. and {Aerts}, C. and {Arenou}, F. and {Cropper}, M. and {Drimmel}, R. and {H{\o}g}, E. and {Katz}, D. and {Lattanzi}, M.~G. and {O'Mullane}, W. and {Grebel}, E.~K. and {Holland}, A.~D. and {Huc}, C. and {Passot}, X. and {Bramante}, L. and {Cacciari}, C. and {Casta{\~n}eda}, J. and {Chaoul}, L. and {Cheek}, N. and {De Angeli}, F. and {Fabricius}, C. and {Guerra}, R. and {Hern{\'a}ndez}, J. and {Jean-Antoine-Piccolo}, A. and {Masana}, E. and {Messineo}, R. and {Mowlavi}, N. and {Nienartowicz}, K. and {Ord{\'o}{\~n}ez-Blanco}, D. and {Panuzzo}, P. and {Portell}, J. and {Richards}, P.~J. and {Riello}, M. and {Seabroke}, G.~M. and {Tanga}, P. and {Th{\'e}venin}, F. and {Torra}, J. and {Els}, S.~G. and {Gracia-Abril}, G. and {Comoretto}, G. and {Garcia-Reinaldos}, M. and {Lock}, T. and {Mercier}, E. and {Altmann}, M. and {Andrae}, R. and {Astraatmadja}, T.~L. and {Bellas-Velidis}, I. and {Benson}, K. and {Berthier}, J. and {Blomme}, R. and {Busso}, G. and {Carry}, B. and {Cellino}, A. and {Clementini}, G. and {Cowell}, S. and {Creevey}, O. and {Cuypers}, J. and {Davidson}, M. and {De Ridder}, J. and {de Torres}, A. and {Delchambre}, L. and {Dell'Oro}, A. and {Ducourant}, C. and {Fr{\'e}mat}, Y. and {Garc{\'\i}a-Torres}, M. and {Gosset}, E. and {Halbwachs}, J. -L. and {Hambly}, N.~C. and {Harrison}, D.~L. and {Hauser}, M. and {Hestroffer}, D. and {Hodgkin}, S.~T. and {Huckle}, H.~E. and {Hutton}, A. and {Jasniewicz}, G. and {Jordan}, S. and {Kontizas}, M. and {Korn}, A.~J. and {Lanzafame}, A.~C. and {Manteiga}, M. and {Moitinho}, A. and {Muinonen}, K. and {Osinde}, J. and {Pancino}, E. and {Pauwels}, T. and {Petit}, J. -M. and {Recio-Blanco}, A. and {Robin}, A.~C. and {Sarro}, L.~M. and {Siopis}, C. and {Smith}, M. and {Smith}, K.~W. and {Sozzetti}, A. and {Thuillot}, W. and {van Reeven}, W. and {Viala}, Y. and {Abbas}, U. and {Abreu Aramburu}, A. and {Accart}, S. and {Aguado}, J.~J. and {Allan}, P.~M. and {Allasia}, W. and {Altavilla}, G. and {{\'A}lvarez}, M.~A. and {Alves}, J. and {Anderson}, R.~I. and {Andrei}, A.~H. and {Anglada Varela}, E. and {Antiche}, E. and {Antoja}, T. and {Ant{\'o}n}, S. and {Arcay}, B. and {Atzei}, A. and {Ayache}, L. and {Bach}, N. and {Baker}, S.~G. and {Balaguer-N{\'u}{\~n}ez}, L. and {Barache}, C. and {Barata}, C. and {Barbier}, A. and {Barblan}, F. and {Baroni}, M. and {Barrado y Navascu{\'e}s}, D. and {Barros}, M. and {Barstow}, M.~A. and {Becciani}, U. and {Bellazzini}, M. and {Bellei}, G. and {Bello Garc{\'\i}a}, A. and {Belokurov}, V. and {Bendjoya}, P. and {Berihuete}, A. and {Bianchi}, L. and {Bienaym{\'e}}, O. and {Billebaud}, F. and {Blagorodnova}, N. and {Blanco-Cuaresma}, S. and {Boch}, T. and {Bombrun}, A. and {Borrachero}, R. and {Bouquillon}, S. and {Bourda}, G. and {Bouy}, H. and {Bragaglia}, A. and {Breddels}, M.~A. and {Brouillet}, N. and {Br{\"u}semeister}, T. and {Bucciarelli}, B. and {Budnik}, F. and {Burgess}, P. and {Burgon}, R. and {Burlacu}, A. and {Busonero}, D. and {Buzzi}, R. and {Caffau}, E. and {Cambras}, J. and {Campbell}, H. and {Cancelliere}, R. and {Cantat-Gaudin}, T. and {Carlucci}, T. and {Carrasco}, J.~M. and {Castellani}, M. and {Charlot}, P. and {Charnas}, J. and {Charvet}, P. and {Chassat}, F. and {Chiavassa}, A. and {Clotet}, M. and {Cocozza}, G. and {Collins}, R.~S. and {Collins}, P. and {Costigan}, G. and {Crifo}, F. and {Cross}, N.~J.~G. and {Crosta}, M. and {Crowley}, C. and {Dafonte}, C. and {Damerdji}, Y. and {Dapergolas}, A. and {David}, P. and {David}, M. and {De Cat}, P. and {de Felice}, F. and {de Laverny}, P. and {De Luise}, F. and {De March}, R. and {de Martino}, D. and {de Souza}, R. and {Debosscher}, J. and {del Pozo}, E. and {Delbo}, M. and {Delgado}, A. and {Delgado}, H.~E. and {di Marco}, F. and {Di Matteo}, P. and {Diakite}, S. and {Distefano}, E. and {Dolding}, C. and {Dos Anjos}, S. and {Drazinos}, P. and {Dur{\'a}n}, J. and {Dzigan}, Y. and {Ecale}, E. and {Edvardsson}, B. and {Enke}, H. and {Erdmann}, M. and {Escolar}, D. and {Espina}, M. and {Evans}, N.~W. and {Eynard Bontemps}, G. and {Fabre}, C. and {Fabrizio}, M. and {Faigler}, S. and {Falc{\~a}o}, A.~J. and {Farr{\`a}s Casas}, M. and {Faye}, F. and {Federici}, L. and {Fedorets}, G. and {Fern{\'a}ndez-Hern{\'a}ndez}, J. and {Fernique}, P. and {Fienga}, A. and {Figueras}, F. and {Filippi}, F. and {Findeisen}, K. and {Fonti}, A. and {Fouesneau}, M. and {Fraile}, E. and {Fraser}, M. and {Fuchs}, J. and {Furnell}, R. and {Gai}, M. and {Galleti}, S. and {Galluccio}, L. and {Garabato}, D. and {Garc{\'\i}a-Sedano}, F. and {Gar{\'e}}, P. and {Garofalo}, A. and {Garralda}, N. and {Gavras}, P. and {Gerssen}, J. and {Geyer}, R. and {Gilmore}, G. and {Girona}, S. and {Giuffrida}, G. and {Gomes}, M. and {Gonz{\'a}lez-Marcos}, A. and {Gonz{\'a}lez-N{\'u}{\~n}ez}, J. and {Gonz{\'a}lez-Vidal}, J.~J. and {Granvik}, M. and {Guerrier}, A. and {Guillout}, P. and {Guiraud}, J. and {G{\'u}rpide}, A. and {Guti{\'e}rrez-S{\'a}nchez}, R. and {Guy}, L.~P. and {Haigron}, R. and {Hatzidimitriou}, D. and {Haywood}, M. and {Heiter}, U. and {Helmi}, A. and {Hobbs}, D. and {Hofmann}, W. and {Holl}, B. and {Holland}, G. and {Hunt}, J.~A.~S. and {Hypki}, A. and {Icardi}, V. and {Irwin}, M. and {Jevardat de Fombelle}, G. and {Jofr{\'e}}, P. and {Jonker}, P.~G. and {Jorissen}, A. and {Julbe}, F. and {Karampelas}, A. and {Kochoska}, A. and {Kohley}, R. and {Kolenberg}, K. and {Kontizas}, E. and {Koposov}, S.~E. and {Kordopatis}, G. and {Koubsky}, P. and {Kowalczyk}, A. and {Krone-Martins}, A. and {Kudryashova}, M. and {Kull}, I. and {Bachchan}, R.~K. and {Lacoste-Seris}, F. and {Lanza}, A.~F. and {Lavigne}, J. -B. and {Le Poncin-Lafitte}, C. and {Lebreton}, Y. and {Lebzelter}, T. and {Leccia}, S. and {Leclerc}, N. and {Lecoeur-Taibi}, I. and {Lemaitre}, V. and {Lenhardt}, H. and {Leroux}, F. and {Liao}, S. and {Licata}, E. and {Lindstr{\o}m}, H.~E.~P. and {Lister}, T.~A. and {Livanou}, E. and {Lobel}, A. and {L{\"o}ffler}, W. and {L{\'o}pez}, M. and {Lopez-Lozano}, A. and {Lorenz}, D. and {Loureiro}, T. and {MacDonald}, I. and {Magalh{\~a}es Fernandes}, T. and {Managau}, S. and {Mann}, R.~G. and {Mantelet}, G. and {Marchal}, O. and {Marchant}, J.~M. and {Marconi}, M. and {Marie}, J. and {Marinoni}, S. and {Marrese}, P.~M. and {Marschalk{\'o}}, G. and {Marshall}, D.~J. and {Mart{\'\i}n-Fleitas}, J.~M. and {Martino}, M. and {Mary}, N. and {Matijevi{\v{c}}}, G. and {Mazeh}, T. and {McMillan}, P.~J. and {Messina}, S. and {Mestre}, A. and {Michalik}, D. and {Millar}, N.~R. and {Miranda}, B.~M.~H. and {Molina}, D. and {Molinaro}, R. and {Molinaro}, M. and {Moln{\'a}r}, L. and {Moniez}, M. and {Montegriffo}, P. and {Monteiro}, D. and {Mor}, R. and {Mora}, A. and {Morbidelli}, R. and {Morel}, T. and {Morgenthaler}, S. and {Morley}, T. and {Morris}, D. and {Mulone}, A.~F. and {Muraveva}, T. and {Musella}, I. and {Narbonne}, J. and {Nelemans}, G. and {Nicastro}, L. and {Noval}, L. and {Ord{\'e}novic}, C. and {Ordieres-Mer{\'e}}, J. and {Osborne}, P. and {Pagani}, C. and {Pagano}, I. and {Pailler}, F. and {Palacin}, H. and {Palaversa}, L. and {Parsons}, P. and {Paulsen}, T. and {Pecoraro}, M. and {Pedrosa}, R. and {Pentik{\"a}inen}, H. and {Pereira}, J. and {Pichon}, B. and {Piersimoni}, A.~M. and {Pineau}, F. -X. and {Plachy}, E. and {Plum}, G. and {Poujoulet}, E. and {Pr{\v{s}}a}, A. and {Pulone}, L. and {Ragaini}, S. and {Rago}, S. and {Rambaux}, N. and {Ramos-Lerate}, M. and {Ranalli}, P. and {Rauw}, G. and {Read}, A. and {Regibo}, S. and {Renk}, F. and {Reyl{\'e}}, C. and {Ribeiro}, R.~A. and {Rimoldini}, L. and {Ripepi}, V. and {Riva}, A. and {Rixon}, G. and {Roelens}, M. and {Romero-G{\'o}mez}, M. and {Rowell}, N. and {Royer}, F. and {Rudolph}, A. and {Ruiz-Dern}, L. and {Sadowski}, G. and {Sagrist{\`a} Sell{\'e}s}, T. and {Sahlmann}, J. and {Salgado}, J. and {Salguero}, E. and {Sarasso}, M. and {Savietto}, H. and {Schnorhk}, A. and {Schultheis}, M. and {Sciacca}, E. and {Segol}, M. and {Segovia}, J.~C. and {Segransan}, D. and {Serpell}, E. and {Shih}, I. -C. and {Smareglia}, R. and {Smart}, R.~L. and {Smith}, C. and {Solano}, E. and {Solitro}, F. and {Sordo}, R. and {Soria Nieto}, S. and {Souchay}, J. and {Spagna}, A. and {Spoto}, F. and {Stampa}, U. and {Steele}, I.~A. and {Steidelm{\"u}ller}, H. and {Stephenson}, C.~A. and {Stoev}, H. and {Suess}, F.~F. and {S{\"u}veges}, M. and {Surdej}, J. and {Szabados}, L. and {Szegedi-Elek}, E. and {Tapiador}, D. and {Taris}, F. and {Tauran}, G. and {Taylor}, M.~B. and {Teixeira}, R. and {Terrett}, D. and {Tingley}, B. and {Trager}, S.~C. and {Turon}, C. and {Ulla}, A. and {Utrilla}, E. and {Valentini}, G. and {van Elteren}, A. and {Van Hemelryck}, E. and {van Leeuwen}, M. and {Varadi}, M. and {Vecchiato}, A. and {Veljanoski}, J. and {Via}, T. and {Vicente}, D. and {Vogt}, S. and {Voss}, H. and {Votruba}, V. and {Voutsinas}, S. and {Walmsley}, G. and {Weiler}, M. and {Weingrill}, K. and {Werner}, D. and {Wevers}, T. and {Whitehead}, G. and {Wyrzykowski}, {\L}. and {Yoldas}, A. and {{\v{Z}}erjal}, M. and {Zucker}, S. and {Zurbach}, C. and {Zwitter}, T. and {Alecu}, A. and {Allen}, M. and {Allende Prieto}, C. and {Amorim}, A. and {Anglada-Escud{\'e}}, G. and {Arsenijevic}, V. and {Azaz}, S. and {Balm}, P. and {Beck}, M. and {Bernstein}, H. -H. and {Bigot}, L. and {Bijaoui}, A. and {Blasco}, C. and {Bonfigli}, M. and {Bono}, G. and {Boudreault}, S. and {Bressan}, A. and {Brown}, S. and {Brunet}, P. -M. and {Bunclark}, P. and {Buonanno}, R. and {Butkevich}, A.~G. and {Carret}, C. and {Carrion}, C. and {Chemin}, L. and {Ch{\'e}reau}, F. and {Corcione}, L. and {Darmigny}, E. and {de Boer}, K.~S. and {de Teodoro}, P. and {de Zeeuw}, P.~T. and {Delle Luche}, C. and {Domingues}, C.~D. and {Dubath}, P. and {Fodor}, F. and {Fr{\'e}zouls}, B. and {Fries}, A. and {Fustes}, D. and {Fyfe}, D. and {Gallardo}, E. and {Gallegos}, J. and {Gardiol}, D. and {Gebran}, M. and {Gomboc}, A. and {G{\'o}mez}, A. and {Grux}, E. and {Gueguen}, A. and {Heyrovsky}, A. and {Hoar}, J. and {Iannicola}, G. and {Isasi Parache}, Y. and {Janotto}, A. -M. and {Joliet}, E. and {Jonckheere}, A. and {Keil}, R. and {Kim}, D. -W. and {Klagyivik}, P. and {Klar}, J. and {Knude}, J. and {Kochukhov}, O. and {Kolka}, I. and {Kos}, J. and {Kutka}, A. and {Lainey}, V. and {LeBouquin}, D. and {Liu}, C. and {Loreggia}, D. and {Makarov}, V.~V. and {Marseille}, M.~G. and {Martayan}, C. and {Martinez-Rubi}, O. and {Massart}, B. and {Meynadier}, F. and {Mignot}, S. and {Munari}, U. and {Nguyen}, A. -T. and {Nordlander}, T. and {Ocvirk}, P. and {O'Flaherty}, K.~S. and {Olias Sanz}, A. and {Ortiz}, P. and {Osorio}, J. and {Oszkiewicz}, D. and {Ouzounis}, A. and {Palmer}, M. and {Park}, P. and {Pasquato}, E. and {Peltzer}, C. and {Peralta}, J. and {P{\'e}turaud}, F. and {Pieniluoma}, T. and {Pigozzi}, E. and {Poels}, J. and {Prat}, G. and {Prod'homme}, T. and {Raison}, F. and {Rebordao}, J.~M. and {Risquez}, D. and {Rocca-Volmerange}, B. and {Rosen}, S. and {Ruiz-Fuertes}, M.~I. and {Russo}, F. and {Sembay}, S. and {Serraller Vizcaino}, I. and {Short}, A. and {Siebert}, A. and {Silva}, H. and {Sinachopoulos}, D. and {Slezak}, E. and {Soffel}, M. and {Sosnowska}, D. and {Strai{\v{z}}ys}, V. and {ter Linden}, M. and {Terrell}, D. and {Theil}, S. and {Tiede}, C. and {Troisi}, L. and {Tsalmantza}, P. and {Tur}, D. and {Vaccari}, M. and {Vachier}, F. and {Valles}, P. and {Van Hamme}, W. and {Veltz}, L. and {Virtanen}, J. and {Wallut}, J. -M. and {Wichmann}, R. and {Wilkinson}, M.~I. and {Ziaeepour}, H. and {Zschocke}, S.},
        title = "{The Gaia mission}",
      journal = {\aap},
         year = 2016,
        month = nov,
       volume = {595},
          eid = {A1},
        pages = {A1},
          doi = {10.1051/0004-6361/201629272},
archivePrefix = {arXiv},
       eprint = {1609.04153},
 primaryClass = {astro-ph.IM},
       adsurl = {https://ui.adsabs.harvard.edu/abs/2016A&A...595A...1G}
}

@ARTICLE{2018A&A...616A...1G,
       author = {{Gaia Collaboration} and {Brown}, A.~G.~A. and {Vallenari}, A. and {Prusti}, T. and {de Bruijne}, J.~H.~J. and {Babusiaux}, C. and {Bailer-Jones}, C.~A.~L. and {Biermann}, M. and {Evans}, D.~W. and {Eyer}, L. and {Jansen}, F. and {Jordi}, C. and {Klioner}, S.~A. and {Lammers}, U. and {Lindegren}, L. and {Luri}, X. and {Mignard}, F. and {Panem}, C. and {Pourbaix}, D. and {Randich}, S. and {Sartoretti}, P. and {Siddiqui}, H.~I. and {Soubiran}, C. and {van Leeuwen}, F. and {Walton}, N.~A. and {Arenou}, F. and {Bastian}, U. and {Cropper}, M. and {Drimmel}, R. and {Katz}, D. and {Lattanzi}, M.~G. and {Bakker}, J. and {Cacciari}, C. and {Casta{\~n}eda}, J. and {Chaoul}, L. and {Cheek}, N. and {De Angeli}, F. and {Fabricius}, C. and {Guerra}, R. and {Holl}, B. and {Masana}, E. and {Messineo}, R. and {Mowlavi}, N. and {Nienartowicz}, K. and {Panuzzo}, P. and {Portell}, J. and {Riello}, M. and {Seabroke}, G.~M. and {Tanga}, P. and {Th{\'e}venin}, F. and {Gracia-Abril}, G. and {Comoretto}, G. and {Garcia-Reinaldos}, M. and {Teyssier}, D. and {Altmann}, M. and {Andrae}, R. and {Audard}, M. and {Bellas-Velidis}, I. and {Benson}, K. and {Berthier}, J. and {Blomme}, R. and {Burgess}, P. and {Busso}, G. and {Carry}, B. and {Cellino}, A. and {Clementini}, G. and {Clotet}, M. and {Creevey}, O. and {Davidson}, M. and {De Ridder}, J. and {Delchambre}, L. and {Dell'Oro}, A. and {Ducourant}, C. and {Fern{\'a}ndez-Hern{\'a}ndez}, J. and {Fouesneau}, M. and {Fr{\'e}mat}, Y. and {Galluccio}, L. and {Garc{\'\i}a-Torres}, M. and {Gonz{\'a}lez-N{\'u}{\~n}ez}, J. and {Gonz{\'a}lez-Vidal}, J.~J. and {Gosset}, E. and {Guy}, L.~P. and {Halbwachs}, J. -L. and {Hambly}, N.~C. and {Harrison}, D.~L. and {Hern{\'a}ndez}, J. and {Hestroffer}, D. and {Hodgkin}, S.~T. and {Hutton}, A. and {Jasniewicz}, G. and {Jean-Antoine-Piccolo}, A. and {Jordan}, S. and {Korn}, A.~J. and {Krone-Martins}, A. and {Lanzafame}, A.~C. and {Lebzelter}, T. and {L{\"o}ffler}, W. and {Manteiga}, M. and {Marrese}, P.~M. and {Mart{\'\i}n-Fleitas}, J.~M. and {Moitinho}, A. and {Mora}, A. and {Muinonen}, K. and {Osinde}, J. and {Pancino}, E. and {Pauwels}, T. and {Petit}, J. -M. and {Recio-Blanco}, A. and {Richards}, P.~J. and {Rimoldini}, L. and {Robin}, A.~C. and {Sarro}, L.~M. and {Siopis}, C. and {Smith}, M. and {Sozzetti}, A. and {S{\"u}veges}, M. and {Torra}, J. and {van Reeven}, W. and {Abbas}, U. and {Abreu Aramburu}, A. and {Accart}, S. and {Aerts}, C. and {Altavilla}, G. and {{\'A}lvarez}, M.~A. and {Alvarez}, R. and {Alves}, J. and {Anderson}, R.~I. and {Andrei}, A.~H. and {Anglada Varela}, E. and {Antiche}, E. and {Antoja}, T. and {Arcay}, B. and {Astraatmadja}, T.~L. and {Bach}, N. and {Baker}, S.~G. and {Balaguer-N{\'u}{\~n}ez}, L. and {Balm}, P. and {Barache}, C. and {Barata}, C. and {Barbato}, D. and {Barblan}, F. and {Barklem}, P.~S. and {Barrado}, D. and {Barros}, M. and {Barstow}, M.~A. and {Bartholom{\'e} Mu{\~n}oz}, S. and {Bassilana}, J. -L. and {Becciani}, U. and {Bellazzini}, M. and {Berihuete}, A. and {Bertone}, S. and {Bianchi}, L. and {Bienaym{\'e}}, O. and {Blanco-Cuaresma}, S. and {Boch}, T. and {Boeche}, C. and {Bombrun}, A. and {Borrachero}, R. and {Bossini}, D. and {Bouquillon}, S. and {Bourda}, G. and {Bragaglia}, A. and {Bramante}, L. and {Breddels}, M.~A. and {Bressan}, A. and {Brouillet}, N. and {Br{\"u}semeister}, T. and {Brugaletta}, E. and {Bucciarelli}, B. and {Burlacu}, A. and {Busonero}, D. and {Butkevich}, A.~G. and {Buzzi}, R. and {Caffau}, E. and {Cancelliere}, R. and {Cannizzaro}, G. and {Cantat-Gaudin}, T. and {Carballo}, R. and {Carlucci}, T. and {Carrasco}, J.~M. and {Casamiquela}, L. and {Castellani}, M. and {Castro-Ginard}, A. and {Charlot}, P. and {Chemin}, L. and {Chiavassa}, A. and {Cocozza}, G. and {Costigan}, G. and {Cowell}, S. and {Crifo}, F. and {Crosta}, M. and {Crowley}, C. and {Cuypers}, J. and {Dafonte}, C. and {Damerdji}, Y. and {Dapergolas}, A. and {David}, P. and {David}, M. and {de Laverny}, P. and {De Luise}, F. and {De March}, R. and {de Martino}, D. and {de Souza}, R. and {de Torres}, A. and {Debosscher}, J. and {del Pozo}, E. and {Delbo}, M. and {Delgado}, A. and {Delgado}, H.~E. and {Di Matteo}, P. and {Diakite}, S. and {Diener}, C. and {Distefano}, E. and {Dolding}, C. and {Drazinos}, P. and {Dur{\'a}n}, J. and {Edvardsson}, B. and {Enke}, H. and {Eriksson}, K. and {Esquej}, P. and {Eynard Bontemps}, G. and {Fabre}, C. and {Fabrizio}, M. and {Faigler}, S. and {Falc{\~a}o}, A.~J. and {Farr{\`a}s Casas}, M. and {Federici}, L. and {Fedorets}, G. and {Fernique}, P. and {Figueras}, F. and {Filippi}, F. and {Findeisen}, K. and {Fonti}, A. and {Fraile}, E. and {Fraser}, M. and {Fr{\'e}zouls}, B. and {Gai}, M. and {Galleti}, S. and {Garabato}, D. and {Garc{\'\i}a-Sedano}, F. and {Garofalo}, A. and {Garralda}, N. and {Gavel}, A. and {Gavras}, P. and {Gerssen}, J. and {Geyer}, R. and {Giacobbe}, P. and {Gilmore}, G. and {Girona}, S. and {Giuffrida}, G. and {Glass}, F. and {Gomes}, M. and {Granvik}, M. and {Gueguen}, A. and {Guerrier}, A. and {Guiraud}, J. and {Guti{\'e}rrez-S{\'a}nchez}, R. and {Haigron}, R. and {Hatzidimitriou}, D. and {Hauser}, M. and {Haywood}, M. and {Heiter}, U. and {Helmi}, A. and {Heu}, J. and {Hilger}, T. and {Hobbs}, D. and {Hofmann}, W. and {Holland}, G. and {Huckle}, H.~E. and {Hypki}, A. and {Icardi}, V. and {Jan{\ss}en}, K. and {Jevardat de Fombelle}, G. and {Jonker}, P.~G. and {Juh{\'a}sz}, {\'A}. L. and {Julbe}, F. and {Karampelas}, A. and {Kewley}, A. and {Klar}, J. and {Kochoska}, A. and {Kohley}, R. and {Kolenberg}, K. and {Kontizas}, M. and {Kontizas}, E. and {Koposov}, S.~E. and {Kordopatis}, G. and {Kostrzewa-Rutkowska}, Z. and {Koubsky}, P. and {Lambert}, S. and {Lanza}, A.~F. and {Lasne}, Y. and {Lavigne}, J. -B. and {Le Fustec}, Y. and {Le Poncin-Lafitte}, C. and {Lebreton}, Y. and {Leccia}, S. and {Leclerc}, N. and {Lecoeur-Taibi}, I. and {Lenhardt}, H. and {Leroux}, F. and {Liao}, S. and {Licata}, E. and {Lindstr{\o}m}, H.~E.~P. and {Lister}, T.~A. and {Livanou}, E. and {Lobel}, A. and {L{\'o}pez}, M. and {Managau}, S. and {Mann}, R.~G. and {Mantelet}, G. and {Marchal}, O. and {Marchant}, J.~M. and {Marconi}, M. and {Marinoni}, S. and {Marschalk{\'o}}, G. and {Marshall}, D.~J. and {Martino}, M. and {Marton}, G. and {Mary}, N. and {Massari}, D. and {Matijevi{\v{c}}}, G. and {Mazeh}, T. and {McMillan}, P.~J. and {Messina}, S. and {Michalik}, D. and {Millar}, N.~R. and {Molina}, D. and {Molinaro}, R. and {Moln{\'a}r}, L. and {Montegriffo}, P. and {Mor}, R. and {Morbidelli}, R. and {Morel}, T. and {Morris}, D. and {Mulone}, A.~F. and {Muraveva}, T. and {Musella}, I. and {Nelemans}, G. and {Nicastro}, L. and {Noval}, L. and {O'Mullane}, W. and {Ord{\'e}novic}, C. and {Ord{\'o}{\~n}ez-Blanco}, D. and {Osborne}, P. and {Pagani}, C. and {Pagano}, I. and {Pailler}, F. and {Palacin}, H. and {Palaversa}, L. and {Panahi}, A. and {Pawlak}, M. and {Piersimoni}, A.~M. and {Pineau}, F. -X. and {Plachy}, E. and {Plum}, G. and {Poggio}, E. and {Poujoulet}, E. and {Pr{\v{s}}a}, A. and {Pulone}, L. and {Racero}, E. and {Ragaini}, S. and {Rambaux}, N. and {Ramos-Lerate}, M. and {Regibo}, S. and {Reyl{\'e}}, C. and {Riclet}, F. and {Ripepi}, V. and {Riva}, A. and {Rivard}, A. and {Rixon}, G. and {Roegiers}, T. and {Roelens}, M. and {Romero-G{\'o}mez}, M. and {Rowell}, N. and {Royer}, F. and {Ruiz-Dern}, L. and {Sadowski}, G. and {Sagrist{\`a} Sell{\'e}s}, T. and {Sahlmann}, J. and {Salgado}, J. and {Salguero}, E. and {Sanna}, N. and {Santana-Ros}, T. and {Sarasso}, M. and {Savietto}, H. and {Schultheis}, M. and {Sciacca}, E. and {Segol}, M. and {Segovia}, J.~C. and {S{\'e}gransan}, D. and {Shih}, I. -C. and {Siltala}, L. and {Silva}, A.~F. and {Smart}, R.~L. and {Smith}, K.~W. and {Solano}, E. and {Solitro}, F. and {Sordo}, R. and {Soria Nieto}, S. and {Souchay}, J. and {Spagna}, A. and {Spoto}, F. and {Stampa}, U. and {Steele}, I.~A. and {Steidelm{\"u}ller}, H. and {Stephenson}, C.~A. and {Stoev}, H. and {Suess}, F.~F. and {Surdej}, J. and {Szabados}, L. and {Szegedi-Elek}, E. and {Tapiador}, D. and {Taris}, F. and {Tauran}, G. and {Taylor}, M.~B. and {Teixeira}, R. and {Terrett}, D. and {Teyssandier}, P. and {Thuillot}, W. and {Titarenko}, A. and {Torra Clotet}, F. and {Turon}, C. and {Ulla}, A. and {Utrilla}, E. and {Uzzi}, S. and {Vaillant}, M. and {Valentini}, G. and {Valette}, V. and {van Elteren}, A. and {Van Hemelryck}, E. and {van Leeuwen}, M. and {Vaschetto}, M. and {Vecchiato}, A. and {Veljanoski}, J. and {Viala}, Y. and {Vicente}, D. and {Vogt}, S. and {von Essen}, C. and {Voss}, H. and {Votruba}, V. and {Voutsinas}, S. and {Walmsley}, G. and {Weiler}, M. and {Wertz}, O. and {Wevers}, T. and {Wyrzykowski}, {\L}. and {Yoldas}, A. and {{\v{Z}}erjal}, M. and {Ziaeepour}, H. and {Zorec}, J. and {Zschocke}, S. and {Zucker}, S. and {Zurbach}, C. and {Zwitter}, T.},
        title = "{Gaia Data Release 2. Summary of the contents and survey properties}",
      journal = {\aap},
         year = 2018,
        month = aug,
       volume = {616},
          eid = {A1},
        pages = {A1},
          doi = {10.1051/0004-6361/201833051},
archivePrefix = {arXiv},
       eprint = {1804.09365},
 primaryClass = {astro-ph.GA},
       adsurl = {https://ui.adsabs.harvard.edu/abs/2018A&A...616A...1G}
}

@ARTICLE{2008AJ....135.2074G,
       author = {{G{\'o}mez}, Jos{\'e} F. and {Su{\'a}rez}, Olga and {G{\'o}mez}, Yolanda and {Miranda}, Luis F. and {Torrelles}, Jos{\'e} M. and {Anglada}, Guillem and {Morata}, {\'O}scar},
        title = "{Radio Interferometric Observations of Candidate Water-Maser-Emitting Planetary Nebulae}",
      journal = {\aj},
         year = 2008,
        month = jun,
       volume = {135},
       number = {6},
        pages = {2074-2083},
          doi = {10.1088/0004-6256/135/6/2074},
archivePrefix = {arXiv},
       eprint = {0803.1644},
 primaryClass = {astro-ph},
       adsurl = {https://ui.adsabs.harvard.edu/abs/2008AJ....135.2074G}
}

@ARTICLE{2015ApJ...799..186G,
       author = {{G{\'o}mez}, Jos{\'e} F. and {Su{\'a}rez}, Olga and {Bendjoya}, Philippe and {Rizzo}, J. Ricardo and {Miranda}, Luis F. and {Green}, James A. and {Uscanga}, Lucero and {Garc{\'\i}a-Garc{\'\i}a}, Enrique and {Lagadec}, Eric and {Guerrero}, Mart{\'\i}n A. and {Ramos-Larios}, Gerardo},
        title = "{The First ``Water Fountain'' Collimated Outflow in a Planetary Nebula}",
      journal = {\apj},
         year = 2015,
        month = feb,
       volume = {799},
       number = {2},
          eid = {186},
        pages = {186},
          doi = {10.1088/0004-637X/799/2/186},
archivePrefix = {arXiv},
       eprint = {1412.2327},
 primaryClass = {astro-ph.SR},
       adsurl = {https://ui.adsabs.harvard.edu/abs/2015ApJ...799..186G}
}

@ARTICLE{2017MNRAS.468.2081G,
       author = {{G{\'o}mez}, Jos{\'e} F. and {Su{\'a}rez}, Olga and {Rizzo}, J. Ricardo and {Uscanga}, Lucero and {Walsh}, Andrew and {Miranda}, Luis F. and {Bendjoya}, Philippe},
        title = "{Interferometric confirmation of `water fountain' candidates}",
      journal = {\mnras},
         year = 2017,
        month = jun,
       volume = {468},
       number = {2},
        pages = {2081-2092},
          doi = {10.1093/mnras/stx650},
archivePrefix = {arXiv},
       eprint = {1703.05037},
 primaryClass = {astro-ph.SR},
       adsurl = {https://ui.adsabs.harvard.edu/abs/2017MNRAS.468.2081G}
}

@MISC{1988ioch.rept.....H,
       author = {{Hanner}, Martha S.},
        title = "{Infrared observations of comets Halley and Wilson and properties of the grains : summary of workshop sponsored by the National Aeronautics and Space Administration, Washington, D.C. and held at Cornell University, Ithaca, New York, August 10-12, 1987}",
 howpublished = {NASA Conference Publication, Summary of a Workshop, held at Cornell University, Ithaca, New York, August 10-12, 1987, Washington: NASA, 1988, edited by Hanner, Martha S.},
         year = 1988,
        month = jan,
       adsurl = {https://ui.adsabs.harvard.edu/abs/1988ioch.rept.....H}
}

@ARTICLE{1983MNRAS.203..301H,
       author = {{Howarth}, I.~D.},
        title = "{LMC and galactic extinction.}",
      journal = {\mnras},
         year = 1983,
        month = apr,
       volume = {203},
        pages = {301-304},
          doi = {10.1093/mnras/203.2.301},
       adsurl = {https://ui.adsabs.harvard.edu/abs/1983MNRAS.203..301H}
}

@ARTICLE{1989ApJ...346..265H,
       author = {{Hrivnak}, Bruce J. and {Kwok}, Sun and {Volk}, Kevin M.},
        title = "{A Study of Several F and G Supergiant-like Stars with Infrared Excesses as Candidates for Proto--Planetary Nebulae}",
      journal = {\apj},
         year = 1989,
        month = nov,
       volume = {346},
        pages = {265},
          doi = {10.1086/168007},
       adsurl = {https://ui.adsabs.harvard.edu/abs/1989ApJ...346..265H}
}

@ARTICLE{2007ApJ...663..342H,
       author = {{Huggins}, P.~J.},
        title = "{Jets and Tori in Proto-Planetary Nebulae}",
      journal = {\apj},
         year = 2007,
        month = jul,
       volume = {663},
       number = {1},
        pages = {342-349},
          doi = {10.1086/518415},
archivePrefix = {arXiv},
       eprint = {astro-ph/0703569},
 primaryClass = {astro-ph},
       adsurl = {https://ui.adsabs.harvard.edu/abs/2007ApJ...663..342H}
}

@INPROCEEDINGS{2007IAUS..242..279I,
       author = {{Imai}, Hiroshi},
        title = "{Stellar molecular jets traced by maser emission}",
    booktitle = {Astrophysical Masers and their Environments},
         year = 2007,
       editor = {{Chapman}, Jessica M. and {Baan}, Willem A.},
       series = {IAU Symposium},
       volume = {242},
        month = mar,
        pages = {279-286},
          doi = {10.1017/S1743921307013130},
archivePrefix = {arXiv},
       eprint = {0709.1797},
 primaryClass = {astro-ph},
       adsurl = {https://ui.adsabs.harvard.edu/abs/2007IAUS..242..279I}
}

@ARTICLE{2010PASJ...62..431I,
       author = {{Imai}, Hiroshi and {Nakashima}, Jun-Ichi and {Deguchi}, Shuji and {Yamauchi}, Aya and {Nakagawa}, Akiharu and {Nagayama}, Takumi},
        title = "{Japanese VLBI Network Mapping of SiO v = 3 J = 1-0 Maser Emission in W Hydrae}",
      journal = {\pasj},
         year = 2010,
        month = apr,
       volume = {62},
        pages = {431},
          doi = {10.1093/pasj/62.2.431},
       adsurl = {https://ui.adsabs.harvard.edu/abs/2010PASJ...62..431I}
}

@ARTICLE{2010A&A...514A...1I,
       author = {{Ishihara}, D. and {Onaka}, T. and {Kataza}, H. and {Salama}, A. and {Alfageme}, C. and {Cassatella}, A. and {Cox}, N. and {Garc{\'\i}a-Lario}, P. and {Stephenson}, C. and {Cohen}, M. and {Fujishiro}, N. and {Fujiwara}, H. and {Hasegawa}, S. and {Ita}, Y. and {Kim}, W. and {Matsuhara}, H. and {Murakami}, H. and {M{\"u}ller}, T.~G. and {Nakagawa}, T. and {Ohyama}, Y. and {Oyabu}, S. and {Pyo}, J. and {Sakon}, I. and {Shibai}, H. and {Takita}, S. and {Tanab{\'e}}, T. and {Uemizu}, K. and {Ueno}, M. and {Usui}, F. and {Wada}, T. and {Watarai}, H. and {Yamamura}, I. and {Yamauchi}, C.},
        title = "{The AKARI/IRC mid-infrared all-sky survey}",
      journal = {\aap},
         year = 2010,
        month = may,
       volume = {514},
          eid = {A1},
        pages = {A1},
          doi = {10.1051/0004-6361/200913811},
archivePrefix = {arXiv},
       eprint = {1003.0270},
 primaryClass = {astro-ph.IM},
       adsurl = {https://ui.adsabs.harvard.edu/abs/2010A&A...514A...1I}
}

@ARTICLE{1999astro.ph.10475I,
       author = {{Ivezic}, Zeljko and {Nenkova}, Maia and {Elitzur}, Moshe},
        title = "{User Manual for DUSTY}",
      journal = {arXiv e-prints},
         year = 1999,
        month = oct,
          eid = {astro-ph/9910475},
        pages = {astro-ph/9910475},
          doi = {10.48550/arXiv.astro-ph/9910475},
archivePrefix = {arXiv},
       eprint = {astro-ph/9910475},
 primaryClass = {astro-ph},
       adsurl = {https://ui.adsabs.harvard.edu/abs/1999astro.ph.10475I}
}

@ARTICLE{2021MNRAS.505.6051J,
       author = {{Jim{\'e}nez-Esteban}, F.~M. and {Engels}, D. and {Aguado}, D.~S. and {Gonz{\'a}lez}, J.~B. and {Garc{\'\i}a-Lario}, P.},
        title = "{An infrared study of Galactic OH/IR stars - III. Variability properties of the Arecibo sample}",
      journal = {\mnras},
         year = 2021,
        month = aug,
       volume = {505},
       number = {4},
        pages = {6051-6068},
          doi = {10.1093/mnras/stab1596},
archivePrefix = {arXiv},
       eprint = {2105.05122},
 primaryClass = {astro-ph.SR},
       adsurl = {https://ui.adsabs.harvard.edu/abs/2021MNRAS.505.6051J}
}

@ARTICLE{2022NatAs...6..275K,
       author = {{Khouri}, Theo and {Vlemmings}, Wouter H.~T. and {Tafoya}, Daniel and {P{\'e}rez-S{\'a}nchez}, Andr{\'e}s F. and {S{\'a}nchez Contreras}, Carmen and {G{\'o}mez}, Jos{\'e} F. and {Imai}, Hiroshi and {Sahai}, Raghvendra},
        title = "{Observational identification of a sample of likely recent common-envelope events}",
      journal = {Nature Astronomy},
         year = 2021,
        month = dec,
       volume = {6},
        pages = {275-286},
          doi = {10.1038/s41550-021-01528-4},
archivePrefix = {arXiv},
       eprint = {2112.09689},
 primaryClass = {astro-ph.SR},
       adsurl = {https://ui.adsabs.harvard.edu/abs/2022NatAs...6..275K}
}

@ARTICLE{1993ARA&A..31...63K,
       author = {{Kwok}, Sun},
        title = "{Proto-planetary nebulae.}",
      journal = {\araa},
         year = 1993,
        month = jan,
       volume = {31},
        pages = {63-92},
          doi = {10.1146/annurev.aa.31.090193.000431},
       adsurl = {https://ui.adsabs.harvard.edu/abs/1993ARA&A..31...63K}
}

@ARTICLE{2007MNRAS.379.1599L,
       author = {{Lawrence}, A. and {Warren}, S.~J. and {Almaini}, O. and {Edge}, A.~C. and {Hambly}, N.~C. and {Jameson}, R.~F. and {Lucas}, P. and {Casali}, M. and {Adamson}, A. and {Dye}, S. and {Emerson}, J.~P. and {Foucaud}, S. and {Hewett}, P. and {Hirst}, P. and {Hodgkin}, S.~T. and {Irwin}, M.~J. and {Lodieu}, N. and {McMahon}, R.~G. and {Simpson}, C. and {Smail}, I. and {Mortlock}, D. and {Folger}, M.},
        title = "{The UKIRT Infrared Deep Sky Survey (UKIDSS)}",
      journal = {\mnras},
         year = 2007,
        month = aug,
       volume = {379},
       number = {4},
        pages = {1599-1617},
          doi = {10.1111/j.1365-2966.2007.12040.x},
archivePrefix = {arXiv},
       eprint = {astro-ph/0604426},
 primaryClass = {astro-ph},
       adsurl = {https://ui.adsabs.harvard.edu/abs/2007MNRAS.379.1599L}
}

@ARTICLE{1977ApJ...217..425M,
       author = {{Mathis}, J.~S. and {Rumpl}, W. and {Nordsieck}, K.~H.},
        title = "{The size distribution of interstellar grains.}",
      journal = {\apj},
         year = 1977,
        month = oct,
       volume = {217},
        pages = {425-433},
          doi = {10.1086/155591},
       adsurl = {https://ui.adsabs.harvard.edu/abs/1977ApJ...217..425M}
}

@INCOLLECTION{1963bad..book..481M,
       author = {{Minkowski}, R.~L. and {Abell}, G.~O.},
        title = "{The National Geographic Society-Palomar Observatory Sky Survey}",
    booktitle = {Basic Astronomical Data: Stars and Stellar Systems},
         year = 1963,
       editor = {{Strand}, K.~A.},
        pages = {481},
       adsurl = {https://ui.adsabs.harvard.edu/abs/1963bad..book..481M}
}

@ARTICLE{2021A&A...649A...1G,
       author = {{Gaia Collaboration} and {Brown}, A.~G.~A. and {Vallenari}, A. and {Prusti}, T. and {de Bruijne}, J.~H.~J. and {Babusiaux}, C. and {Biermann}, M. and {Creevey}, O.~L. and {Evans}, D.~W. and {Eyer}, L. and {Hutton}, A. and {Jansen}, F. and {Jordi}, C. and {Klioner}, S.~A. and {Lammers}, U. and {Lindegren}, L. and {Luri}, X. and {Mignard}, F. and {Panem}, C. and {Pourbaix}, D. and {Randich}, S. and {Sartoretti}, P. and {Soubiran}, C. and {Walton}, N.~A. and {Arenou}, F. and {Bailer-Jones}, C.~A.~L. and {Bastian}, U. and {Cropper}, M. and {Drimmel}, R. and {Katz}, D. and {Lattanzi}, M.~G. and {van Leeuwen}, F. and {Bakker}, J. and {Cacciari}, C. and {Casta{\~n}eda}, J. and {De Angeli}, F. and {Ducourant}, C. and {Fabricius}, C. and {Fouesneau}, M. and {Fr{\'e}mat}, Y. and {Guerra}, R. and {Guerrier}, A. and {Guiraud}, J. and {Jean-Antoine Piccolo}, A. and {Masana}, E. and {Messineo}, R. and {Mowlavi}, N. and {Nicolas}, C. and {Nienartowicz}, K. and {Pailler}, F. and {Panuzzo}, P. and {Riclet}, F. and {Roux}, W. and {Seabroke}, G.~M. and {Sordo}, R. and {Tanga}, P. and {Th{\'e}venin}, F. and {Gracia-Abril}, G. and {Portell}, J. and {Teyssier}, D. and {Altmann}, M. and {Andrae}, R. and {Bellas-Velidis}, I. and {Benson}, K. and {Berthier}, J. and {Blomme}, R. and {Brugaletta}, E. and {Burgess}, P.~W. and {Busso}, G. and {Carry}, B. and {Cellino}, A. and {Cheek}, N. and {Clementini}, G. and {Damerdji}, Y. and {Davidson}, M. and {Delchambre}, L. and {Dell'Oro}, A. and {Fern{\'a}ndez-Hern{\'a}ndez}, J. and {Galluccio}, L. and {Garc{\'\i}a-Lario}, P. and {Garcia-Reinaldos}, M. and {Gonz{\'a}lez-N{\'u}{\~n}ez}, J. and {Gosset}, E. and {Haigron}, R. and {Halbwachs}, J.-L. and {Hambly}, N.~C. and {Harrison}, D.~L. and {Hatzidimitriou}, D. and {Heiter}, U. and {Hern{\'a}ndez}, J. and {Hestroffer}, D. and {Hodgkin}, S.~T. and {Holl}, B. and {Jan{\ss}en}, K. and {Jevardat de Fombelle}, G. and {Jordan}, S. and {Krone-Martins}, A. and {Lanzafame}, A.~C. and {L{\"o}ffler}, W. and {Lorca}, A. and {Manteiga}, M. and {Marchal}, O. and {Marrese}, P.~M. and {Moitinho}, A. and {Mora}, A. and {Muinonen}, K. and {Osborne}, P. and {Pancino}, E. and {Pauwels}, T. and {Petit}, J.-M. and {Recio-Blanco}, A. and {Richards}, P.~J. and {Riello}, M. and {Rimoldini}, L. and {Robin}, A.~C. and {Roegiers}, T. and {Rybizki}, J. and {Sarro}, L.~M. and {Siopis}, C. and {Smith}, M. and {Sozzetti}, A. and {Ulla}, A. and {Utrilla}, E. and {van Leeuwen}, M. and {van Reeven}, W. and {Abbas}, U. and {Abreu Aramburu}, A. and {Accart}, S. and {Aerts}, C. and {Aguado}, J.~J. and {Ajaj}, M. and {Altavilla}, G. and {{\'A}lvarez}, M.~A. and {{\'A}lvarez Cid-Fuentes}, J. and {Alves}, J. and {Anderson}, R.~I. and {Anglada Varela}, E. and {Antoja}, T. and {Audard}, M. and {Baines}, D. and {Baker}, S.~G. and {Balaguer-N{\'u}{\~n}ez}, L. and {Balbinot}, E. and {Balog}, Z. and {Barache}, C. and {Barbato}, D. and {Barros}, M. and {Barstow}, M.~A. and {Bartolom{\'e}}, S. and {Bassilana}, J.-L. and {Bauchet}, N. and {Baudesson-Stella}, A. and {Becciani}, U. and {Bellazzini}, M. and {Bernet}, M. and {Bertone}, S. and {Bianchi}, L. and {Blanco-Cuaresma}, S. and {Boch}, T. and {Bombrun}, A. and {Bossini}, D. and {Bouquillon}, S. and {Bragaglia}, A. and {Bramante}, L. and {Breedt}, E. and {Bressan}, A. and {Brouillet}, N. and {Bucciarelli}, B. and {Burlacu}, A. and {Busonero}, D. and {Butkevich}, A.~G. and {Buzzi}, R. and {Caffau}, E. and {Cancelliere}, R. and {C{\'a}novas}, H. and {Cantat-Gaudin}, T. and {Carballo}, R. and {Carlucci}, T. and {Carnerero}, M.~I. and {Carrasco}, J.~M. and {Casamiquela}, L. and {Castellani}, M. and {Castro-Ginard}, A. and {Castro Sampol}, P. and {Chaoul}, L. and {Charlot}, P. and {Chemin}, L. and {Chiavassa}, A. and {Cioni}, M.-R.~L. and {Comoretto}, G. and {Cooper}, W.~J. and {Cornez}, T. and {Cowell}, S. and {Crifo}, F. and {Crosta}, M. and {Crowley}, C. and {Dafonte}, C. and {Dapergolas}, A. and {David}, M. and {David}, P.},
        title = "{Gaia Early Data Release 3. Summary of the contents and survey properties}",
      journal = {\aap},
         year = 2021,
        month = may,
       volume = {649},
          eid = {A1},
        pages = {A1},
          doi = {10.1051/0004-6361/202039657},
archivePrefix = {arXiv},
       eprint = {2012.01533},
 primaryClass = {astro-ph.GA},
       adsurl = {https://ui.adsabs.harvard.edu/abs/2021A&A...649A...1G}
}

@ARTICLE{2014RMxAA..50..293M,
       author = {{Molina}, R.~E. and {Giridhar}, S. and {Pereira}, C.~B. and {Arellano Ferro}, A. and {Muneer}, S.},
        title = "{Spectroscopic analysis of four post-AGB candidates}",
      journal = {\rmxaa},
         year = 2014,
        month = oct,
       volume = {50},
        pages = {293-306},
          doi = {10.48550/arXiv.1405.6746},
archivePrefix = {arXiv},
       eprint = {1405.6746},
 primaryClass = {astro-ph.SR},
       adsurl = {https://ui.adsabs.harvard.edu/abs/2014RMxAA..50..293M}
}

@ARTICLE{2009ApJ...692..402N,
       author = {{Nakashima}, Jun-ichi and {Koning}, Nico and {Kwok}, Sun and {Zhang}, Yong},
        title = "{Morphokinematic Properties of the 21 {\ensuremath{\mu}}m Source IRAS 07134+1005}",
      journal = {\apj},
         year = 2009,
        month = feb,
       volume = {692},
       number = {1},
        pages = {402-410},
          doi = {10.1088/0004-637X/692/1/402},
archivePrefix = {arXiv},
       eprint = {0810.4383},
 primaryClass = {astro-ph},
       adsurl = {https://ui.adsabs.harvard.edu/abs/2009ApJ...692..402N}
}

@ARTICLE{2012ApJ...759...61N,
       author = {{Nakashima}, Jun-ichi and {Koning}, Nico and {Volgenau}, Nikolaus H. and {Kwok}, Sun and {Yung}, Bosco H.~K. and {Zhang}, Yong},
        title = "{CO Structure of the 21 {\ensuremath{\mu}}m Source IRAS 22272+5435: A Sign of a Jet Launch?}",
      journal = {\apj},
         year = 2012,
        month = nov,
       volume = {759},
       number = {1},
          eid = {61},
        pages = {61},
          doi = {10.1088/0004-637X/759/1/61},
archivePrefix = {arXiv},
       eprint = {1209.4168},
 primaryClass = {astro-ph.SR},
       adsurl = {https://ui.adsabs.harvard.edu/abs/2012ApJ...759...61N}
}

@ARTICLE{1998A&AS..130....1N,
       author = {{Neri}, R. and {Kahane}, C. and {Lucas}, R. and {Bujarrabal}, V. and {Loup}, C.},
        title = "{A (12) CO (J=1-> 0) and (J=2-> 1) atlas of circumstellar envelopes of AGB and post-AGB stars}",
      journal = {\aaps},
         year = 1998,
        month = may,
       volume = {130},
        pages = {1-64},
          doi = {10.1051/aas:1998213},
       adsurl = {https://ui.adsabs.harvard.edu/abs/1998A&AS..130....1N}
}

@ARTICLE{1984ApJ...278L...1N,
       author = {{Neugebauer}, G. and {Habing}, H.~J. and {van Duinen}, R. and {Aumann}, H.~H. and {Baud}, B. and {Beichman}, C.~A. and {Beintema}, D.~A. and {Boggess}, N. and {Clegg}, P.~E. and {de Jong}, T. and {Emerson}, J.~P. and {Gautier}, T.~N. and {Gillett}, F.~C. and {Harris}, S. and {Hauser}, M.~G. and {Houck}, J.~R. and {Jennings}, R.~E. and {Low}, F.~J. and {Marsden}, P.~L. and {Miley}, G. and {Olnon}, F.~M. and {Pottasch}, S.~R. and {Raimond}, E. and {Rowan-Robinson}, M. and {Soifer}, B.~T. and {Walker}, R.~G. and {Wesselius}, P.~R. and {Young}, E.},
        title = "{The Infrared Astronomical Satellite (IRAS) mission.}",
      journal = {\apjl},
         year = 1984,
        month = mar,
       volume = {278},
        pages = {L1-L6},
          doi = {10.1086/184209},
       adsurl = {https://ui.adsabs.harvard.edu/abs/1984ApJ...278L...1N}
}

@ARTICLE{1998A&AS..127..185N,
       author = {{Nyman}, L. -A. and {Hall}, P.~J. and {Olofsson}, H.},
        title = "{SiO masers in OH/IR stars, proto-planetary and planetary nebulae}",
      journal = {\aaps},
         year = 1998,
        month = jan,
       volume = {127},
        pages = {185-200},
          doi = {10.1051/aas:1998343},
       adsurl = {https://ui.adsabs.harvard.edu/abs/1998A&AS..127..185N}
}

@ARTICLE{2017AJ....153..119O,
       author = {{Orosz}, G. and {Imai}, H. and {Dodson}, R. and {Rioja}, M.~J. and {Frey}, S. and {Burns}, R.~A. and {Etoka}, S. and {Nakagawa}, A. and {Nakanishi}, H. and {Asaki}, Y. and {Goldman}, S.~R. and {Tafoya}, D.},
        title = "{Astrometry of OH/IR Stars Using 1612 MHz Hydroxyl Masers. I. Annual Parallaxes of WX Psc and OH138.0+7.2}",
      journal = {\aj},
         year = 2017,
        month = mar,
       volume = {153},
       number = {3},
          eid = {119},
        pages = {119},
          doi = {10.3847/1538-3881/153/3/119},
archivePrefix = {arXiv},
       eprint = {1701.05101},
 primaryClass = {astro-ph.SR},
       adsurl = {https://ui.adsabs.harvard.edu/abs/2017AJ....153..119O}
}

@ARTICLE{1992A&A...261..567O,
       author = {{Ossenkopf}, V. and {Henning}, Th. and {Mathis}, J.~S.},
        title = "{Constraints on cosmic silicates.}",
      journal = {\aap},
         year = 1992,
        month = aug,
       volume = {261},
        pages = {567-578},
       adsurl = {https://ui.adsabs.harvard.edu/abs/1992A&A...261..567O}
}

@ARTICLE{1988A&A...194..335P,
       author = {{Pegourie}, B.},
        title = "{Optical properties of alpha silicon carbide.}",
      journal = {\aap},
         year = 1988,
        month = apr,
       volume = {194},
        pages = {335-339},
       adsurl = {https://ui.adsabs.harvard.edu/abs/1988A&A...194..335P}
}

@ARTICLE{1990ApJ...360L..51R,
       author = {{Reid}, M.~J. and {Menten}, K.~M.},
        title = "{A Subarcsecond H 2O Maser Shell Surrounding a Variable Star}",
      journal = {\apjl},
         year = 1990,
        month = sep,
       volume = {360},
        pages = {L51},
          doi = {10.1086/185810},
       adsurl = {https://ui.adsabs.harvard.edu/abs/1990ApJ...360L..51R}
}

@ARTICLE{2011A&A...525A..56R,
       author = {{Richards}, A.~M.~S. and {Elitzur}, M. and {Yates}, J.~A.},
        title = "{Observational evidence for the shrinking of bright maser spots}",
      journal = {\aap},
         year = 2011,
        month = jan,
       volume = {525},
          eid = {A56},
        pages = {A56},
          doi = {10.1051/0004-6361/201015397},
archivePrefix = {arXiv},
       eprint = {1010.4419},
 primaryClass = {astro-ph.GA},
       adsurl = {https://ui.adsabs.harvard.edu/abs/2011A&A...525A..56R}
}

@ARTICLE{2011ApJ...737..103S,
       author = {{Schlafly}, Edward F. and {Finkbeiner}, Douglas P.},
        title = "{Measuring Reddening with Sloan Digital Sky Survey Stellar Spectra and Recalibrating SFD}",
      journal = {\apj},
         year = 2011,
        month = aug,
       volume = {737},
       number = {2},
          eid = {103},
        pages = {103},
          doi = {10.1088/0004-637X/737/2/103},
archivePrefix = {arXiv},
       eprint = {1012.4804},
 primaryClass = {astro-ph.GA},
       adsurl = {https://ui.adsabs.harvard.edu/abs/2011ApJ...737..103S}
}

@ARTICLE{2002AJ....123.2772S,
       author = {{Sevenster}, Maartje N.},
        title = "{OH-selected AGB and Post-AGB Objects. I. Infrared and Maser Properties}",
      journal = {\aj},
         year = 2002,
        month = may,
       volume = {123},
       number = {5},
        pages = {2772-2787},
          doi = {10.1086/339827},
archivePrefix = {arXiv},
       eprint = {astro-ph/0202182},
 primaryClass = {astro-ph},
       adsurl = {https://ui.adsabs.harvard.edu/abs/2002AJ....123.2772S}
}

@ARTICLE{2002AJ....123.2788S,
       author = {{Sevenster}, Maartje N.},
        title = "{OH-selected AGB and Post-AGB Objects. II. Blue versus Red Evolution off the Asymptotic Giant Branch}",
      journal = {\aj},
         year = 2002,
        month = may,
       volume = {123},
       number = {5},
        pages = {2788-2795},
          doi = {10.1086/339828},
archivePrefix = {arXiv},
       eprint = {astro-ph/0202183},
 primaryClass = {astro-ph},
       adsurl = {https://ui.adsabs.harvard.edu/abs/2002AJ....123.2788S}
}

@ARTICLE{2006AJ....131.1163S,
       author = {{Skrutskie}, M.~F. and {Cutri}, R.~M. and {Stiening}, R. and {Weinberg}, M.~D. and {Schneider}, S. and {Carpenter}, J.~M. and {Beichman}, C. and {Capps}, R. and {Chester}, T. and {Elias}, J. and {Huchra}, J. and {Liebert}, J. and {Lonsdale}, C. and {Monet}, D.~G. and {Price}, S. and {Seitzer}, P. and {Jarrett}, T. and {Kirkpatrick}, J.~D. and {Gizis}, J.~E. and {Howard}, E. and {Evans}, T. and {Fowler}, J. and {Fullmer}, L. and {Hurt}, R. and {Light}, R. and {Kopan}, E.~L. and {Marsh}, K.~A. and {McCallon}, H.~L. and {Tam}, R. and {Van Dyk}, S. and {Wheelock}, S.},
        title = "{The Two Micron All Sky Survey (2MASS)}",
      journal = {\aj},
         year = 2006,
        month = feb,
       volume = {131},
       number = {2},
        pages = {1163-1183},
          doi = {10.1086/498708},
       adsurl = {https://ui.adsabs.harvard.edu/abs/2006AJ....131.1163S}
}

@ARTICLE{2000MNRAS.312..217S,
       author = {{Soker}, Noam},
        title = "{Dust formation and inhomogeneous mass-loss from asymptotic giant branch stars}",
      journal = {\mnras},
         year = 2000,
        month = feb,
       volume = {312},
       number = {1},
        pages = {217-224},
          doi = {10.1046/j.1365-8711.2000.03156.x},
archivePrefix = {arXiv},
       eprint = {astro-ph/9908320},
 primaryClass = {astro-ph},
       adsurl = {https://ui.adsabs.harvard.edu/abs/2000MNRAS.312..217S}
}

@ARTICLE{2008ApJ...689..430S,
       author = {{Su{\'a}rez}, Olga and {G{\'o}mez}, Jos{\'e} F. and {Miranda}, Luis F.},
        title = "{VLA Observations of the ``Water Fountain'' IRAS 16552-3050}",
      journal = {\apj},
         year = 2008,
        month = dec,
       volume = {689},
       number = {1},
        pages = {430-435},
          doi = {10.1086/592493},
archivePrefix = {arXiv},
       eprint = {0809.0392},
 primaryClass = {astro-ph},
       adsurl = {https://ui.adsabs.harvard.edu/abs/2008ApJ...689..430S}
}

@dataset{1995yCat.2161....0S,
       author = {{Sweeney}, L.~H. and {Richardson}, T.},
        title = "{VizieR Online Data Catalog: Equatorial Infrared Catalog (Sweeney+ 1990)}",
 howpublished = {VizieR On-line Data Catalog: II/161.  Originally published in: Space Applications Corporation 1990},
         year = 1995,
        month = may,
          eid = {II/161},
       adsurl = {https://ui.adsabs.harvard.edu/abs/1995yCat.2161....0S}
}

@ARTICLE{2019A&A...629A...8T,
       author = {{Tafoya}, D. and {Orosz}, G. and {Vlemmings}, W.~H.~T. and {Sahai}, R. and {P{\'e}rez-S{\'a}nchez}, A.~F.},
        title = "{Spatio-kinematical model of the collimated molecular outflow in the water-fountain nebula IRAS 16342-3814}",
      journal = {\aap},
         year = 2019,
        month = sep,
       volume = {629},
          eid = {A8},
        pages = {A8},
          doi = {10.1051/0004-6361/201834632},
archivePrefix = {arXiv},
       eprint = {1906.06328},
 primaryClass = {astro-ph.SR},
       adsurl = {https://ui.adsabs.harvard.edu/abs/2019A&A...629A...8T}
}

@ARTICLE{2001ApJ...557..831U,
       author = {{Ueta}, Toshiya and {Meixner}, Margaret and {Hinz}, Philip M. and {Hoffmann}, William F. and {Brandner}, Wolfgang and {Dayal}, Aditya and {Deutsch}, Lynne K. and {Fazio}, Giovanni G. and {Hora}, Joseph L.},
        title = "{Subarcsecond Mid-Infrared Structure of the Dust Shell around IRAS 22272+5435}",
      journal = {\apj},
         year = 2001,
        month = aug,
       volume = {557},
       number = {2},
        pages = {831-843},
          doi = {10.1086/322259},
archivePrefix = {arXiv},
       eprint = {astro-ph/0104437},
 primaryClass = {astro-ph},
       adsurl = {https://ui.adsabs.harvard.edu/abs/2001ApJ...557..831U}
}

@ARTICLE{2012A&A...547A..40U,
       author = {{Uscanga}, L. and {G{\'o}mez}, J.~F. and {Su{\'a}rez}, O. and {Miranda}, L.~F.},
        title = "{An updated catalog of OH-maser-emitting planetary nebulae}",
      journal = {\aap},
         year = 2012,
        month = nov,
       volume = {547},
          eid = {A40},
        pages = {A40},
          doi = {10.1051/0004-6361/201219760},
archivePrefix = {arXiv},
       eprint = {1209.5768},
 primaryClass = {astro-ph.SR},
       adsurl = {https://ui.adsabs.harvard.edu/abs/2012A&A...547A..40U}
}

@ARTICLE{1988A&A...194..125V,
       author = {{van der Veen}, W.~E.~C.~J. and {Habing}, H.~J.},
        title = "{The IRAS two-colour diagram as a tool for studying late stages of stellar evolution.}",
      journal = {\aap},
         year = 1988,
        month = apr,
       volume = {194},
        pages = {125-134},
       adsurl = {https://ui.adsabs.harvard.edu/abs/1988A&A...194..125V}
}

@ARTICLE{2016AJ....152...16V,
       author = {{van Belle}, Gerard T. and {Creech-Eakman}, Michelle J. and {Ruiz-Velasco}, Alma E.},
        title = "{Bolometric Flux Estimation for Cool Evolved Stars}",
      journal = {\aj},
         year = 2016,
        month = jul,
       volume = {152},
       number = {1},
          eid = {16},
        pages = {16},
          doi = {10.3847/0004-6256/152/1/16},
archivePrefix = {arXiv},
       eprint = {1604.00984},
 primaryClass = {astro-ph.SR},
       adsurl = {https://ui.adsabs.harvard.edu/abs/2016AJ....152...16V}
}

@ARTICLE{1993ApJ...413..641V,
       author = {{Vassiliadis}, E. and {Wood}, P.~R.},
        title = "{Evolution of Low- and Intermediate-Mass Stars to the End of the Asymptotic Giant Branch with Mass Loss}",
      journal = {\apj},
         year = 1993,
        month = aug,
       volume = {413},
        pages = {641},
          doi = {10.1086/173033},
       adsurl = {https://ui.adsabs.harvard.edu/abs/1993ApJ...413..641V}
}

@ARTICLE{2003ARA&A..41..391V,
       author = {{van Winckel}, Hans},
        title = "{Post-AGB Stars}",
      journal = {\araa},
         year = 2003,
        month = jan,
       volume = {41},
        pages = {391-427},
          doi = {10.1146/annurev.astro.41.071601.170018},
       adsurl = {https://ui.adsabs.harvard.edu/abs/2003ARA&A..41..391V}
}

@ARTICLE{2014AA...569A..92V,
       author = {{Vlemmings}, W.~H.~T. and {Amiri}, N. and {van Langevelde}, H.~J. and {Tafoya}, D.},
        title = "{From the ashes: JVLA observations of water fountain nebula candidates show the rebirth of IRAS 18455+0448}",
      journal = {\aap},
         year = 2014,
        month = sep,
       volume = {569},
          eid = {A92},
        pages = {A92},
          doi = {10.1051/0004-6361/201423754},
archivePrefix = {arXiv},
       eprint = {1407.6709},
 primaryClass = {astro-ph.SR},
       adsurl = {https://ui.adsabs.harvard.edu/abs/2014A&A...569A..92V}
}

@INPROCEEDINGS{1994AAS...185.4515W,
       author = {{Winfrey}, S. and {Barnbaum}, C. and {Morris}, M. and {Omont}, A.},
        title = "{Spectral Classification of Cold IRAS Stars: Supergiants}",
    booktitle = {American Astronomical Society Meeting Abstracts},
         year = 1994,
       series = {American Astronomical Society Meeting Abstracts},
       volume = {185},
        month = dec,
          eid = {45.15},
        pages = {45.15},
       adsurl = {https://ui.adsabs.harvard.edu/abs/1994AAS...185.4515W}
}

@ARTICLE{1992ApJ...397..552W,
       author = {{Wood}, P.~R. and {Whiteoak}, J.~B. and {Hughes}, S.~M.~G. and {Bessell}, M.~S. and {Gardner}, F.~F. and {Hyland}, A.~R.},
        title = "{OH/IR Stars in the Magellanic Clouds}",
      journal = {\apj},
         year = 1992,
        month = oct,
       volume = {397},
        pages = {552},
          doi = {10.1086/171812},
       adsurl = {https://ui.adsabs.harvard.edu/abs/1992ApJ...397..552W}
}

@ARTICLE{2012A&A...537A...5W,
       author = {{Wolak}, P. and {Szymczak}, M. and {G{\'e}rard}, E.},
        title = "{Polarization properties of OH masers in AGB and post-AGB stars}",
      journal = {\aap},
         year = 2012,
        month = jan,
       volume = {537},
          eid = {A5},
        pages = {A5},
          doi = {10.1051/0004-6361/201117263},
archivePrefix = {arXiv},
       eprint = {1110.0773},
 primaryClass = {astro-ph.GA},
       adsurl = {https://ui.adsabs.harvard.edu/abs/2012A&A...537A...5W}
}
